\documentclass[10pt,aps,pra,twocolumn,nofootinbib]{revtex4-1}   

\usepackage[dvips]{epsfig}
\usepackage{amsmath}    
\usepackage{amsfonts}  
\usepackage{amssymb}
\usepackage{graphicx}   
\usepackage{mathrsfs}
\usepackage{ulem}
\usepackage{soul}
\usepackage[usenames,dvipsnames]{color}

\usepackage{xcolor}

\usepackage[colorlinks]{hyperref}

\usepackage{colortbl}

\definecolor{gray(x11gray)}{rgb}{0.75, 0.75, 0.75}

\definecolor{darksienna}{rgb}{0.24, 0.08, 0.08}

\definecolor{darkblue}{rgb}{0.0, 0.13, 0.35}

\newcommand{\bra}[1]{\langle #1 |}
\newcommand{\ket}[1]{| #1 \rangle}

\newcommand{\ignore}[1]{}

\newcommand{\blue}{\color{blue}}

\AtBeginDocument{%
    \hypersetup{
      urlcolor     = blue, 
      filecolor   = red,
      linkcolor    = blue, 
      citecolor   = blue, 
    }
} 

\begin{document}

\title{Single- versus two-parameter Fisher information in quantum interferometry}

\author{Stefan~Ataman}
\affiliation{Extreme Light Infrastructure - Nuclear Physics (ELI-NP), `Horia Hulubei' National R\&D Institute for Physics and Nuclear Engineering (IFIN-HH), 30 Reactorului Street, 077125 M\u{a}gurele, jud. Ilfov, Romania}
\email{stefan.ataman@eli-np.ro}

\date{\today}


\begin{abstract}
In this paper we reconsider the single parameter quantum Fisher information (QFI) and compare it with the two-parameter one. We find simple relations connecting the single parameter QFI (both in the asymmetric and symmetric phase shift cases) to the two parameter Fisher matrix coefficients. Following some clarifications about the role of an external phase [Phys. Rev. A \textbf{85}, 011801(R) (2012)], the single-parameter QFI and its over-optimistic predictions have been disregarded in the literature. We show in this paper that both the  single- and two-parameter QFI have physical meaning and their predicted quantum Cram\'er-Rao bounds are often attainable with the appropriate experimental setup. Moreover, we give practical situations of interest in quantum metrology, where the phase sensitivities of a number of input states approach the quantum Cram\'er-Rao bound induced by the single-parameter QFI, outperforming the two-parameter QFI.

\end{abstract}

\maketitle

\section{Introduction}
\label{sec:introduction}

Interferometric phase sensitivity is a research topic of interest for a number of rapidly growing scientific fields, among which we can single out gravitational wave astronomy \cite{LIGO13,Gro13,Ace14,Oel14,Meh18,Vah18,Tse2019,Dem13} and quantum technologies \cite{Gio12,Dow15,Pez18,Xu19}.

With the advent of non-classical states of light \cite{Yue76,Yur85,Lou87,Hof07}, the classical SNL (shot-noise limit) \cite{Mic18} has been shown to be improvable \cite{Cav81}, prediction confirmed by experiments \cite{Xia87,Hol93,Kuz98}. 

Theoretical bounds for the interferometric phase sensitivity became possible due to the quantum Fisher information (QFI) and its associated quantum Cram\'er-Rao bound (QCRB) \cite{Bra94,Dem12,Dem15,Pez15,Par09,Pez09}. These bounds, besides their theoretical interest, are extremely useful in evaluating the optimality of realistic detection schemes.


Jarzyna \& Demkowicz-Dobrza\ifmmode \acute{n}\else \'{n}\fi{}ski \cite{Jar12} showed in a convincing manner that using the single-parameter QFI constantly yields over-optimistic results. As pointed out, the solution to avoid counting resources that are actually unavailable is to phase-average the input state \cite{Jar12,Zho17,Tak17} or use the two-parameter QFI \cite{Lan13,Lan14,API18,Pre19,Wei20,Ata19}.


One reason that the single parameter QFI might be considered over-optimistic or artificial is that usual detection schemes cannot go beyond the QCRB given by a two-parameter QFI approach \cite{Gar17,API18,Lan13}. Even the balanced homodyne detection, although having access to an external phase reference, cannot exceed this limit if the interferometer is balanced \cite{Gar17,Ata19}. 

As discussed in previous works \cite{Jar12,Tak17}, actual phase measurement scenarios are modeled with a single phase shift for some applications \cite{Tay13,Ono13}, while others require two phase shifts \cite{LIGO13}. Thus, we consider both these scenarios in this work.


Gaussian input states are a popular choice due to both their properties and to technical advancements in their preparation \cite{And16}. Among them we can cite the coherent plus squeezed vacuum input state \cite{Cav81,Pez07,Bar03}, a popular choice also due to its use in gravitational wave detection \cite{Ace14,Oel14,Meh18,Vah18,Tse2019}. The squeezed coherent plus squeezed vacuum input state \cite{Par95} has been shown to bring a gain in phase sensitivity due to the second squeezer \cite{Pre19,Ata19}. This gain, however, becomes marginal in the experimentally interesting scenario of high input coherent power and limited squeezing factors. In this paper, we will show how to overcome this limitation using an unbalanced interferometer and an external  phase reference. We also note that, on the experimental side, squeezing a laser source has been recently demonstrated \cite{Vah18}.

Although most authors employ balanced (50/50) interferometers  \cite{Gar17,Dem15,Pez07,Pez08,Bar03}, a number of works  addressed the unbalanced scenarios, too \cite{Jar12,Pre19,Wei20}. Some interesting results emerged, for example in the case of double coherent input \cite{Pre19}.

Reference \cite{Jar12} gave reasons not to use a single-parameter QFI. In this paper we take exactly the opposite route: we find scenarios where using a single-parameter QFI is interesting. Moreover, we find detection schemes that are actually able to reach the QCRB predicted by the single-parameter QFI. However, in order to do so, we need to employ an unbalanced interferometer.


The input phase matching conditions (PMC) and their effect on performance have been discussed in the literature \cite{Jar12,Liu13,Pre19,Ata19}. Although in some works all input phases are set to zero \cite{Jar12,Liu13,Par95}, this is not always an optimal choice \cite{Pre19,Ata19}. In this paper we will show that the optimal PMCs change not only in function of the input state, but also with the type of QFI used.


In this work we focus on two detection schemes. The difference-intensity detection scheme is often considered in the literature \cite{Dem15,Gar17,API18,Pre19,Ata19} and it is a good example of a detection method not having access to an external phase reference. We thus expect its performance to be limited by the two-parameter QFI. The homodyne detection
technique \cite{Yue78,Vog83,Gen12,Gar17,Ata19} is the quintessential example of a detector having access to an external phase reference. We will show that under the right conditions, it is able outperform the QCRB implied by the two-parameter QFI, approaching the one corresponding to the single parameter QFI.


This paper is structured as follows. In Section \ref{sec:Fisher_information_2param} we introduce some conventions and describe the two-parameter QFI approach. In Section \ref{sec:Fisher_information_1param_asymetric} we discuss the single-parameter QFI with an asymmetric phase shift while in Section \ref{sec:Fisher_information_1param_symetric_varphi_over2} we discuss the same problem in the symmetric phase shift scenario. In Section \ref{sec:Fisher_input_Gaussian_states} we give the complete expression for these QFIs for an important class of input states, namely the Gaussian states. The realistic detection schemes to be considered in this paper are described in Section \ref{sec:realistic_detection_schemes}. The performance of these schemes with some Gaussian input states is detailed and discussed in Section \ref{sec:phase_sensitivity_gaussian_states}. The results are discussed and some assertions from the literature commented in Section \ref{sec:discussion}. The paper closes  conclusions in Section \ref{sec:conclusions}.

\section{Two parameter quantum Fisher information}
\label{sec:Fisher_information_2param}

Throughout this work we assume no losses and our input is limited to a pure state, thus we do not need to use the Symmetric Logarithmic Derivative \cite{Dem15,Bra94,Par09}. We also assume no entanglement between the two input ports. This is a rather standard assumption in papers discussing Gaussian input states \cite{Ata19,Pre19,API18,Lan13,Lan14,Gar17}.

\begin{figure}
\includegraphics[scale=0.75]{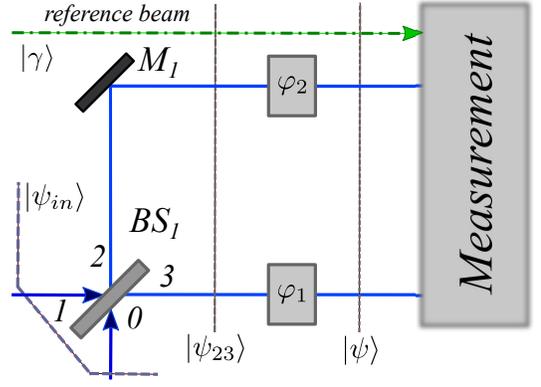}
\caption{The configuration for the case study with two independent phase shifts, $\varphi_1$ and $\varphi_2$. The beam splitter $BS_1$ is assumed to have a variable transmission coefficient, $T$.}
\label{fig:MZI_Fisher_two_phases}
\end{figure}

We first consider the general case where each arm of the interferometer contains a phase-shift ($\varphi_1$ and, respectively, $\varphi_2$, see Fig.~\ref{fig:MZI_Fisher_two_phases}). BS denotes the beam splitter. The estimation is treated as a general two parameter problem \cite{Lan13,Lan14,API18,Jar12}. We define the $2\times2$ Fisher information matrix \cite{Par09,Lan13,Lan14}
\begin{equation}
\label{eq:Fisher_matrix}
   \mathcal{F}=
  \left[ {\begin{array}{cc}
  \mathcal{F}_{ss} & \mathcal{F}_{sd} \\
   \mathcal{F}_{ds} & \mathcal{F}_{dd} \\
  \end{array} } \right]
\end{equation}
where the coefficients are defined by
\begin{equation}
\label{eq:Fisher_matrix_elements}
\mathcal{F}_{ij}=4\Re\{\langle\partial_i\psi\vert\partial_j\psi\rangle-\langle\partial_i\psi\vert\psi\rangle
 \langle\psi\vert\partial_j\psi\rangle\}
\end{equation}
with $i, j\in \{s,d\}$, ${\varphi_{s/d}=\varphi_1\pm\varphi_2}$ and $\Re$ denotes the real part. We also denote ${\vert\partial_{s/d}\psi\rangle=\partial\vert\psi\rangle/\partial{\varphi_{s/d}}}$ and we have ${\vert\psi\rangle=e^{-i\frac{\hat{n}_2-\hat{n}_3}{2}\varphi_d}e^{-i\frac{\hat{n}_2+\hat{n}_3}{2}\varphi_s}\vert\psi_{23}\rangle}$ where $\hat{n}_m=\hat{a}_m^\dagger\hat{a}_m$ denotes the number operator for port (mode) $m$. We employ the usual annihilation (creation) operators $\hat{a}_m$ ($\hat{a}_m^\dagger$) obeying the commutation relations $[\hat{a}_m,\hat{a}_n^\dagger]=\delta_{mn}$ with $m,n$ labeling spatial modes.

From the Fisher matrix \eqref{eq:Fisher_matrix} we arrive at a QCRB matrix inequality \cite{Lan13} out of which we retain only the difference-difference phase estimator,
\begin{equation}
\label{eq:Delta_varphi_geq_Fisher_2p_dd}
(\Delta\varphi_d)^2\geq(\mathcal{F}^{-1})_{dd}
\end{equation}
and since the matrix element $(\mathcal{F}^{-1})_{dd}$ will appear repeatedly we define the two-parameter QFI,
\begin{eqnarray}
\label{eq:F_2p_definition}
\mathcal{F}^{(2p)}:=\frac{1}{(\mathcal{F}^{-1})_{dd}}
=\mathcal{F}_{dd}-\frac{\mathcal{F}_{sd}\mathcal{F}_{ds}}{\mathcal{F}_{ss}}
\end{eqnarray}
thus saturating inequality \eqref{eq:Delta_varphi_geq_Fisher_2p_dd} implies the two-parameter QCRB,
\begin{equation}
\label{eq:Delta_varphi_QCRB_2p_DEFINTION}
\Delta\varphi^{(2p)}_{QCRB}=\frac{1}{\sqrt{\mathcal{F}^{(2p)}}}.
\end{equation}
Since the QFI is additive (both for the single- and two parameter cases) \cite{Bra94,Dem15}, for $N$ repeated experiments we have the scaling $\Delta\varphi^{(2p)}_{QCRB}=\sqrt{N\mathcal{F}^{(2p)}}$. For simplicity, throughout this paper we set $N=1$.

For the calculation of the Fisher matrix elements we need the field operator transformations,
\begin{equation}
\label{eq:field_op_transf_MZI_a_hat}
\left\{
\begin{array}{l}
\hat{a}_3=R\hat{a}_0+T\hat{a}_1\\
\hat{a}_2=T\hat{a}_0+R\hat{a}_1
\end{array}
\right.
\end{equation}
where $T$ ($R$) denotes the transmission (reflection) coefficient of the beam splitter ($BS_1$ in Fig.~\ref{fig:MZI_Fisher_two_phases}). We have $\vert{T}\vert^2+\vert{R}\vert^2=1$ and $TR^*+T^*R=0$ \cite{GerryKnight}. Since the last relation implies $(T^*R)^2=-|TR|^2$, a sign convention has to be made (i. e. $T^*R=\pm i|TR|$). Without loss of generality, throughout this paper we use the convention $iT^*R=-\vert{TR}\vert$ and for the particular case of balanced BS we consider $T=1/\sqrt{2}$ and $R=i/\sqrt{2}$.

Using the definition from equation \eqref{eq:Fisher_matrix_elements}, the sum-sum Fisher matrix element $\mathcal{F}_{ss}$ can now be computed and yields
\begin{eqnarray}
\label{eq:app:F_ss_FINAL_FORM_GENERAL}
\mathcal{F}_{ss}
=\Delta^2\hat{n}_0+\Delta^2\hat{n}_1.
\end{eqnarray}
By the variance $\Delta^2\hat{n}_k$ we denote $\bra{\psi}\hat{n}_k^2\ket{\psi}-\bra{\psi}\hat{n}_k\ket{\psi}^2$ and the standard deviation is $\Delta\hat{n}_k=\sqrt{\Delta^2\hat{n}_k}$. $\mathcal{F}_{dd}$ is computed and the result is given in equation \eqref{eq:F_dd_FINAL_FORM_GENERAL_compact}. The last term we need is $\mathcal{F}_{sd}$ since $\mathcal{F}_{sd}=\mathcal{F}_{ds}$ \cite{Lan13} and the result is given in equation \eqref{eq:F_sd_GENERIC_FINAL_compact}.

In the balanced case $\mathcal{F}_{ss}$ remains unchanged while the difference-difference Fisher matrix element $\mathcal{F}_{dd}$ becomes
\begin{eqnarray}
\label{eq:app:F_dd_FINAL_FORM_GENERAL_balanced}
\mathcal{F}_{dd}=
\langle\hat{n}_1\rangle+\langle\hat{n}_0\rangle
+2\left(
\langle\hat{n}_0\rangle\langle\hat{n}_1\rangle
-\vert\langle\hat{a}_0\rangle\vert^2\vert\langle\hat{a}_1\rangle\vert^2\right)
\nonumber\\
-2\Re\left(
\langle\hat{a}_0^2\rangle\langle(\hat{a}_1^\dagger)^2\rangle
-\langle\hat{a}_0\rangle^2\langle\hat{a}_1^\dagger\rangle^2
\right)
\quad
\end{eqnarray}
and $\mathcal{F}_{sd}$ reduces to
\begin{eqnarray}
\label{eq:app:F_sd_GENERIC_FINAL_balanced}
\mathcal{F}_{sd}
=2\Im\left(\langle\hat{a}_0\rangle\langle\hat{a}_1^\dagger\rangle
+\left(\langle\hat{n}_0\hat{a}_0\rangle
-\langle\hat{n}_0\rangle\langle\hat{a}_0\rangle\right)\langle\hat{a}_1^\dagger\rangle
\right.
\nonumber\\
\left.
+\langle\hat{a}_0\rangle
\left(\langle\hat{a}_1^\dagger\hat{n}_1\rangle
-\langle\hat{a}_1^\dagger\rangle
\langle\hat{n}_1\rangle\right)
\right)
\end{eqnarray}
where $\Im$ denotes the imaginary part \cite{Ata19}. Throughout this work we assume that the input port $1$ is never in the vacuum state, i. e. $\langle\hat{n}_1\rangle\neq0$.


Interferometric  phase sensitivity is based on the phase difference induced between the two arms of an interferometer. In most cases the optimal sensitivity is obtained in the balanced case \cite{Lan13,Lan14}, but exceptions have been shown to exist \cite{Tak17,Pre19}. Thus, if we stray away from the balanced case until the extreme $|T|\to1$ (or $|T|\to0$), one can assume that there can be no interferometric phase sensitivity (except when using an external phase reference).

We can quickly estimate the predictions of the extreme $|T|\to1$ case. Obviously $\mathcal{F}_{ss}$ from equation \eqref{eq:app:F_ss_FINAL_FORM_GENERAL} remains unchained and applying the limit $|T|\to1$ to equation \eqref{eq:F_dd_FINAL_FORM_GENERAL_compact} yields $\mathcal{F}_{dd}=\Delta^2\hat{n}_0+\Delta^2\hat{n}_1$. The same constraint applied to $\mathcal{F}_{sd}$ from equation \eqref{eq:F_sd_GENERIC_FINAL_compact} yields
\begin{eqnarray}
\label{eq:F_sd_GENERIC_FINAL_compact_T_equal_0}
\mathcal{F}_{sd}
=\Delta^2{\hat{n}_1}
-\Delta^2{\hat{n}_0}.
\end{eqnarray}
If $\mathcal{F}_{sd}=0$ i. e. if $\Delta^2\hat{n}_0=\Delta^2\hat{n}_1$ then equation \eqref{eq:F_2p_definition} implies
\begin{equation}
\label{eq:F_2p_T_equal_0_F_sd_is_zero}
\mathcal{F}^{(2p)}=2\Delta^2\hat{n}_1.
\end{equation}
If $\mathcal{F}_{sd}\neq0$, the two-parameter difference-difference equivalent QFI becomes
\begin{equation}
\label{eq:F_2p_T_equal_0_F_sd_not_zero}
\mathcal{F}^{(2p)}
=4\frac{\Delta^2\hat{n}_1\Delta^2\hat{n}_0}
{\Delta^2\hat{n}_0
+\Delta^2\hat{n}_1}
\end{equation}
and somehow surprisingly $\mathcal{F}^{(2p)}\neq0$ except when the input state $0$ is in the vacuum state. However, one should not overlook the fact that although the two-parameter QFI guarantees not to consider resources obtainable via an external phase reference, the input being in a pure state, it implies a fixed phase relation between the quantum states from ports $0$ and $1$. Since a well chosen detection scheme can take advantage of this fact, equation \eqref{eq:F_2p_T_equal_0_F_sd_not_zero} should be less surprising.

\section{Single parameter quantum Fisher information with an asymmetric phase shift}
\label{sec:Fisher_information_1param_asymetric}
\begin{figure}
\includegraphics[scale=0.75]{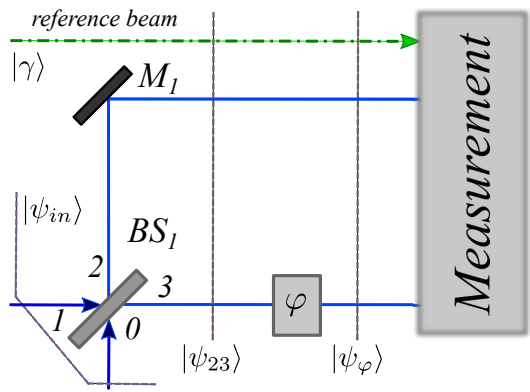}
\caption{The configuration for the case study employing a single phase shift. The beam splitter $BS_1$ is assumed to have a variable transmission coefficient, $T$. }
\label{fig:MZI_Fisher_single_phase}
\end{figure}

In the single-parameter case (see Fig.~\ref{fig:MZI_Fisher_single_phase}), the QFI is simply \cite{Par09,Dem15,Bra94}
\begin{equation}
\label{eq:Fisher_information_single_i}
\mathcal{F}^{(i)} = 4\left(\langle\partial_\varphi\psi\vert\partial_\varphi\psi\rangle-\vert\langle\partial_\varphi\psi\vert\psi\rangle\vert^2\right),
\end{equation}
and we use here the notations from reference \cite{Jar12}. We assume a single phase shift in the output $3$ of $BS_1$, i. e. we model it as  $\hat{U}(\varphi)=e^{-i\varphi\hat{n}_3}$, thus the QFI is formally given by  
\begin{equation}
\mathcal{F}^{(i)}=4\Delta^2\hat{n}_3
\end{equation}
and it implies the (single parameter) QCRB
\begin{equation}
\label{eq:Delta_varphi_QCRB_i_DEFINTION}
\Delta\varphi^{(i)}_{QCRB}=\frac{1}{\sqrt{\mathcal{F}^{(i)}}}.
\end{equation}
The calculations for $\mathcal{F}^{(i)}$ are detailed in Appendix \ref{sec:app:single_Fisher_information_asym} and
the final result with respect to the input parameters is given in equation \eqref{eq:Fisher_sg_param_nonBal_symBS_separable_input_GENERAL}. 
%
%
Comparing equation \eqref{eq:Fisher_sg_param_nonBal_symBS_separable_input_GENERAL} with the Fisher matrix elements from the previous section we note that the single-parameter Fisher information can be expressed as a function of the Fisher matrix elements and we have
\begin{equation}
\label{eq:F_i_is_Fss_plus_Fdd-2Fds}
\mathcal{F}^{(i)}=\mathcal{F}_{ss}+\mathcal{F}_{dd}-2\mathcal{F}_{sd}.
\end{equation}
From the definition of the two-parameter QFI \eqref{eq:F_2p_definition} and the above equation we can immediately prove that
\begin{equation}
\label{eq:F_i_is_BIGGER_or_EQUAL_than_F2p}
\mathcal{F}^{(i)}\geq\mathcal{F}^{(2p)}
\end{equation}
with equality only if $\mathcal{F}_{ss}=\mathcal{F}_{sd}$. In the balanced case the QFI $\mathcal{F}^{(i)}$ simplifies to the expression given by equation \eqref{eq:Fisher_sg_param_nonBal_symBS_separable_input_balanced}. In the limit case $|T|\to1$, the QFI from equation \eqref{eq:Fisher_sg_param_nonBal_symBS_separable_input_GENERAL} reduces to
\begin{equation}
\mathcal{F}^{(i)}=4\Delta^2\hat{n}_1
\end{equation}
a result that might look surprising, since all terms related to input port $0$ are missing. However, in this degenerate case only input port $1$, can ``reach'' the phase shift $\varphi$, hence the result.

\section{Single parameter quantum Fisher information with two symmetric phase shifts}
\label{sec:Fisher_information_1param_symetric_varphi_over2}

\begin{figure}
\includegraphics[scale=0.75]{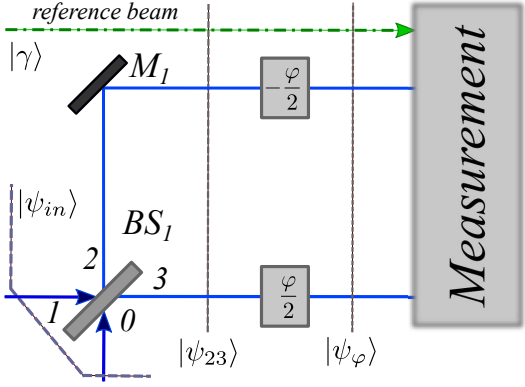}
\caption{The configuration for the case study with symmetric, (anti)correlated phase shifts, $\pm\varphi/2$. The beam splitter $BS_1$ is assumed to have a variable transmission coefficient, $T$.}
\label{fig:MZi_Fisher_single_param_sym_varphi_over_2}
\end{figure}

In this last scenario (see Fig.~ \ref{fig:MZi_Fisher_single_param_sym_varphi_over_2}), we assume a distributed phase shift of $\varphi/2$ ($-\varphi/2$) in the output port $3$ ($2$) of the beam splitter $BS_1$, i. e. we model it as $\hat{U}(\varphi)=e^{-i\frac{\varphi}{2}\hat{n}_3+i\frac{\varphi}{2}\hat{n}_2}$, thus the QFI is given by  
\begin{equation}
\label{eq:Fisher_information_single_ii}
\mathcal{F}^{(ii)}=\Delta^2{\hat{n}_2}+\Delta^2{\hat{n}_3}.
\end{equation}
The calculations are detailed in Appendix \ref{sec:app:single_Fisher_information_sym} and the final form of $\mathcal{F}^{(ii)}$ is given in equation \eqref{eq:Fisher_information_sg_param_symmetrical_varphi_over_2}. This QFI implies the QCRB
\begin{equation}
\label{eq:Delta_varphi_QCRB_ii_DEFINTION}
\Delta\varphi_{QCRB}^{(ii)}=\frac{1}{\sqrt{\mathcal{F}^{(ii)}}}.
\end{equation}
%
In this scenario, too, we find a simple relation connecting $\mathcal{F}^{(ii)}$ to the two-parameter Fisher matrix elements, namely
\begin{equation}
\mathcal{F}^{(ii)}=\frac{\mathcal{F}_{ss}}{2}+\frac{\mathcal{F}_{dd}}{2}
\end{equation}
This time, however, there is no relation of type \eqref{eq:F_i_is_BIGGER_or_EQUAL_than_F2p} between $\mathcal{F}^{(ii)}$ and $\mathcal{F}^{(2p)}$.

In the balanced case $\mathcal{F}^{(ii)}$ simplifies to the expression given by equation \eqref{eq:Fisher_information_sg_param_symmetrical_varphi_over_2_BALANCED}. In the limit $|T|\to1$ we have 
\begin{equation}
\mathcal{F}^{(ii)}=\Delta^2\hat{n}_0+\Delta^2\hat{n}_1.
\end{equation}

\section{Gaussian input states and their respective QFI}
\label{sec:Fisher_input_Gaussian_states}
In this section we discuss the three previously introduced QFI metrics (i. e. $\mathcal{F}^{(2p)}$, $\mathcal{F}^{(i)}$ and $\mathcal{F}^{(ii)}$) with a number of Gaussian input states.

\subsection{Single coherent input}
\label{subsec:Fisher_input_Gaussian_states_single_coh}
In this simple scenario we consider the input state 
\begin{equation}
\label{eq:psi_in_single_coh}
\vert\psi_{in}\rangle=\vert\alpha_1\rangle=\hat{D}_1\left(\alpha\right)\ket{0}
\end{equation}
where the displacement or Glauber operator \cite{GerryKnight,MandelWolf,Aga12} for a port $k$ is defined by 
\begin{equation}
\label{eq:displacement_operator_def}
\hat{D}_k\left(\alpha\right)=e^{\alpha\hat{a}_k^\dagger-\alpha^*\hat{a}_k}.
\end{equation}
The first Fisher matrix element is $\mathcal{F}_{ss}=\vert\alpha\vert^2$ and from equation \eqref{eq:F_dd_FINAL_FORM_GENERAL_compact} we get
\begin{eqnarray}
\mathcal{F}_{dd}=\left(\vert{T}\vert^2-\vert{R}\vert^2\right)^2\vert\alpha\vert^2
+4\vert{TR}\vert^2\vert\alpha\vert^2
=\vert\alpha\vert^2
\end{eqnarray}
Finally, equation \eqref{eq:F_sd_GENERIC_FINAL_compact} gives $\mathcal{F}_{sd}=-\left(\vert{T}\vert^2-\vert{R}\vert^2\right)\vert\alpha\vert^2$ and from equation \eqref{eq:F_2p_definition} we obtain the two-parameter QFI,
\begin{equation}
\label{eq:Fisher_2p_single_coherent}
\mathcal{F}^{(2p)}
=4\vert{TR}\vert^2\vert\alpha\vert^2
\end{equation}
and it implies the QCRB
\begin{equation}
\label{eq:Delta_varphi_QCRB_2p_single_coherent}
\Delta\varphi_{QCRB}^{(2p)}=\frac{1}{\sqrt{\mathcal{F}^{(2p)}}}
=\frac{1}{2\vert{TR}\vert\vert\alpha\vert}
\end{equation}
yielding in the balanced case the well-known shot-noise limit $\Delta\varphi_{QCRB}^{(2p)}=1/\vert\alpha\vert$ \cite{Dem15,Lan13,Gar17}. This limit has been shown to be achieved with difference-intensity \cite{Dem15,Gar17,API18} single-mode intensity \cite{Gar17,API18} as well as balanced homodyne detection schemes \cite{Gar17,Ata19}.

The single-parameter QFI from equation \eqref{eq:Fisher_information_single_i} is found to be
\begin{equation}
\label{eq:Fisher_i_single_coherent}
\mathcal{F}^{(i)}=4\vert{T}\vert^2\vert\alpha\vert^2
\end{equation}
implying a QCRB
\begin{equation}
\label{eq:Delta_varphi_QCRB_i_single_coherent}
\Delta\varphi_{QCRB}^{(i)}
=\frac{1}{{2}\vert{T}\vert\vert\alpha\vert}.
\end{equation}
For the balanced case we get $\mathcal{F}^{(i)}=2\vert\alpha\vert^2$, thus we already improve the phases sensitivity, ${\Delta\varphi^{(i)}_{QCRB}=1/\sqrt{2}\vert\alpha\vert}$. We can even go further and consider the ``unphysical'' case $|T|\to1$ yielding $\Delta\varphi^{(i)}_{QCRB}=1/2\vert\alpha\vert$. In Section \ref{subsec:phase_sensitivity_gaussian_states_single_coh} we will show that there is nothing unphysical about this scenario, it all depends on how we intend to measure our phase sensitivity.

\begin{figure}
\centering
\includegraphics[scale=0.45]{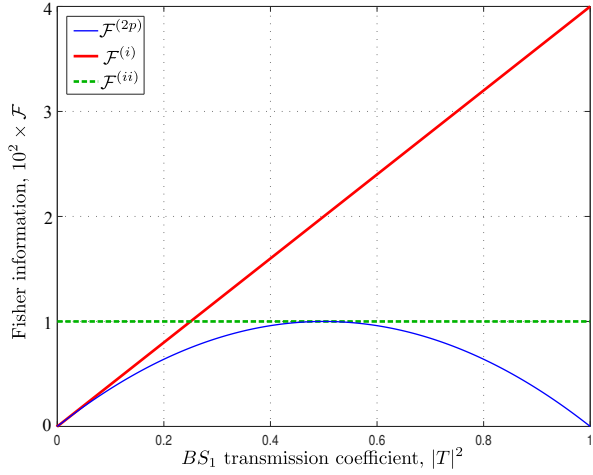}
\caption{The three QFIs versus the transmission coefficient of $BS_1$ for a single coherent input state with $\vert\alpha\vert=10$. While $\mathcal{F}^{(ii)}$ remains constant, irrespective of $|T|^2$, $\mathcal{F}^{(i)}$ steadily grows from $0$ to $4\vert\alpha\vert^2$. The two parameter QFI reaches its optimum in the balanced case (i. e. $|T|^2=0.5$).}
\label{fig:Fisher_2p_i_ii_single_coh}
\end{figure}

In the symmetric phase shift case, equation \eqref{eq:Fisher_information_single_ii} gives
\begin{equation}
\label{eq:Fisher_ii_single_coherent}
\mathcal{F}^{(ii)}=\vert\alpha\vert^2
\end{equation}
with the corresponding QCRB,
\begin{equation}
\label{eq:Delta_varphi_QCRB_ii_single_coherent}
\Delta\varphi_{QCRB}^{(ii)}
=\frac{1}{\vert\alpha\vert}.
\end{equation}
In Fig.~\ref{fig:Fisher_2p_i_ii_single_coh} we plot the three discussed QFIs versus $|T|^2$ for $\vert\alpha\vert=10$. With $\mathcal{F}^{(ii)}$ remaining constant, regardless of the value of $T$, the ``true'' phase sensitivity $\mathcal{F}^{(2p)}$ peaks for a balanced beam splitter yielding the well-known result $\mathcal{F}^{(2p)}=\vert\alpha\vert^2$ \cite{Dem15,Jar12} while $\mathcal{F}^{(i)}$ steadily grows reaching its maximum value $\mathcal{F}^{(i)}_{max}=4\vert\alpha\vert^2$ for $|T|=1$.

\subsection{Double coherent input}
\label{subsec:Fisher_input_Gaussian_states_dual_coh}
In this scenario we consider the input state
\begin{equation}
\label{eq:psi_in_dual_coh}
\vert\psi_{in}\rangle=\vert\alpha_1\beta_0\rangle=\hat{D}_1\left(\alpha\right)\hat{D}_0\left(\beta\right)\vert0\rangle
\end{equation}
where we denote $\alpha=\vert\alpha\vert e^{i\theta_\alpha}$, $\beta=\vert\alpha\vert e^{i\theta_\beta}$ and  $\Delta\theta= \theta_\alpha-\theta_\beta$. The first Fisher matrix element yields $\mathcal{F}_{ss}=\vert\alpha\vert^2+\vert\beta\vert^2$. From equation \eqref{eq:F_dd_FINAL_FORM_GENERAL_compact} we have
\begin{equation}
\label{eq:F_dd_dual_coh}
\mathcal{F}_{dd}=\vert\alpha\vert^2+\vert\beta\vert^2
\end{equation}
and the last Fisher matrix element yields
\begin{equation}
\label{eq:F_sd_dual_coh}
\mathcal{F}_{sd}=\left(\vert{T}\vert^2-\vert{R}\vert^2\right)(\vert\beta\vert^2-\vert\alpha\vert^2)-4|TR|\vert\alpha\beta\vert\sin\Delta\theta.
\end{equation}
Using these results we get two-parameter QFI and its expression is given in equation \eqref{eq:Fisher_2p_dual_coherent}. As proved in reference \cite{Pre19}, for $\vert\alpha\vert$, $\vert\beta\vert$ and $\Delta\theta$ given, an optimum transmission coefficient exists and it is given by
\begin{equation}
\label{eq:T_opt_squared_versus_Delta_theta_dual_coh_F_2p}
{T}_{opt}^{(2p)}=
\sqrt{\frac{1}{2}+
\frac{\mathrm{sign}(\varpi^2-1)\varpi \sin\Delta\theta}
{\sqrt{\left(1-\varpi^2\right)^2
+4\varpi^2\sin^2\Delta\theta}
}}
\end{equation}
where $\varpi=\vert\beta\vert/\vert\alpha\vert$. Replacing $|T|$ with ${T}_{opt}^{(2p)}$ in equation \eqref{eq:Fisher_2p_dual_coherent} (and assuming ${T}_{opt}^{(2p)}\neq\{0,1\}$) brings $\mathcal{F}^{(2p)}$ to its global maximum,
\begin{eqnarray}
\label{eq:Fisher_2p_dual_coherent_PMC_optimal}
\mathcal{F}^{(2p)}_{max}
=\vert\alpha\vert^2
+\vert\beta\vert^2
\end{eqnarray}
implying the QCRB
\begin{equation}
\label{eq:QCRB_F_2p_dual_coh}
\Delta\varphi_{QCRB}^{(2p)}=\frac{1}{\sqrt{\vert\alpha\vert^2+\vert\beta\vert^2}}.
\end{equation}
In the asymmetric single phase shift case (see Fig.~\ref{fig:MZI_Fisher_single_phase}) the single-parameter QFI equation \eqref{eq:Fisher_information_single_i} yields
\begin{equation}
\label{eq:Fisher_i_double_coherent}
\mathcal{F}^{(i)}=4|T|^2\vert\alpha\vert^2
+4|R|^2\vert\beta\vert^2
+8|TR|\vert\alpha\beta\vert\sin\Delta\theta
\end{equation}
While for the two-parameter QFI, regardless of the value of the input PMC ($\Delta\theta$) we can find an optimum transmission coefficient \eqref{eq:T_opt_squared_versus_Delta_theta_dual_coh_F_2p} bringing us to the maximal QFI \eqref{eq:Fisher_2p_dual_coherent_PMC_optimal}, this is no longer true for $\mathcal{F}^{(i)}$.  If  $\sin\Delta\theta=0$, then $\mathcal{F}^{(i)}$ is maximized for $|T|=1$ (if $\vert\alpha\vert>\vert\beta\vert$) or for $T=0$ (if $\vert\alpha\vert<\vert\beta\vert$). If $\sin\Delta\theta\neq0$, the optimal transmission coefficient is given by equation \eqref{eq:T_opt_squared_versus_Delta_theta_dual_coh_F_i}. For $\sin\Delta\theta=1$  equation \eqref{eq:T_opt_squared_versus_Delta_theta_dual_coh_F_i} yields the simple expression
\begin{equation}
\label{eq:T_opt_Delta_theta_pi_over_2_dual_coh_F_i}
T_{opt}^{(i)}
=\frac{|\alpha|}{\sqrt{|\alpha|^2+|\beta|^2}}.
\end{equation}
Replacing this result into equation \eqref{eq:Fisher_i_double_coherent} takes $\mathcal{F}^{(i)}$ to its global maximum
\begin{equation}
\label{eq:Fisher_i_double_coherent_MAX}
\mathcal{F}^{(i)}_{max}
=4(\vert\alpha\vert^2+\vert\beta\vert^2)
\end{equation}
and this implies the QCRB
\begin{equation}
\label{eq:QCRB_F_i_dual_coh}
\Delta\varphi_{QCRB}^{(i)}=\frac{1}{2\sqrt{\vert\alpha\vert^2+\vert\beta\vert^2}}.
\end{equation}

\begin{figure}
\centering
\includegraphics[scale=0.45]{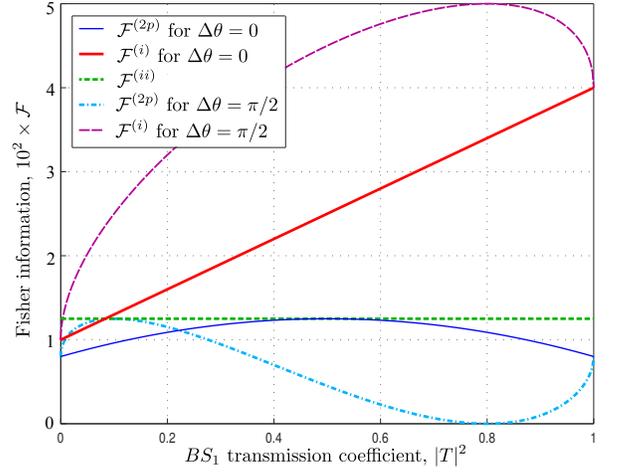}
\caption{The three QFIs versus the transmission coefficient of $BS_1$ for a double coherent input state, with two phase-matching conditions. While for $\Delta\theta=0$, $\mathcal{F}^{(i)}$ linearly grows from $4\vert\beta\vert^2$ to $4\vert\alpha\vert^2$, for $\Delta\theta=\pi/2$ it peaks at $4(\vert\alpha\vert^2+\vert\beta\vert^2)$ for $|T|={T}^{(i)}_{{opt}}$. Parameters used: $\vert\alpha\vert=10$, $\vert\beta\vert=5$.}
\label{fig:Fisher_2p_i_ii_double_coh}
\end{figure}

\noindent Finally, in the symmetrical case (see Fig.~\ref{fig:MZi_Fisher_single_param_sym_varphi_over_2}), from equation \eqref{eq:Fisher_information_single_ii} we get
\begin{equation}
\label{eq:Fisher_ii_dual_coherent}
\mathcal{F}^{(ii)}=\vert\alpha\vert^2+\vert\beta\vert^2
\end{equation}
and it implies $\Delta\varphi_{QCRB}^{(ii)}=1/\sqrt{\vert\alpha\vert^2+\vert\beta\vert^2}$. Remarkably, this QFI is totally immune to the input PMC and to the transmission coefficient of $BS_1$.

In Fig.~\ref{fig:Fisher_2p_i_ii_double_coh} we plot the three QFI metrics against the transmission coefficient of $BS_1$, $|T|^2$. We consider $\vert\alpha\vert>\vert\beta\vert$ and we first discuss the case $\Delta\theta=0$. While $\mathcal{F}^{(ii)}$ remains constant irrespective of the values taken by $T$ and $\Delta\theta$, $\mathcal{F}^{(i)}$ varies linearly from $4\vert\beta\vert^2$ (for $|T|=0$) to $4\vert\alpha\vert^2$ (for $|T|=1$). The two-parameter QFI attains its maximum value in the balanced case. In the extreme case $|T|=0/1$, regardless of the value of $\Delta\theta$, it reaches $\mathcal{F}^{(2p)}=4\vert\alpha\beta\vert^2/(\vert\alpha\vert^2+\vert\beta\vert^2)$ in agreement with equation \eqref{eq:F_2p_T_equal_0_F_sd_not_zero}. For $\Delta\theta\neq0$, the behavior of both $\mathcal{F}^{(2p)}$ and $\mathcal{F}^{(i)}$ changes.

In the case of a two-parameter QFI, while the maximum attainable QFI \eqref{eq:Fisher_2p_dual_coherent_PMC_optimal} remains unchanged, the optimum transmission coefficient shifts from the balanced case \cite{Pre19}. For $\Delta\theta=\pi/2$ and the values given in Fig.~\ref{fig:Fisher_2p_i_ii_double_coh}, one finds  ${T}^{(2p)}_{opt}\approx\sqrt{0.13}$. The asymmetric single-parameter QFI is maximized to $\mathcal{F}^{(i)}_{max}=4(\vert\alpha\vert^2+\vert\beta\vert^2)$ for the transmission coefficient $T_{opt}^{(i)}\approx\sqrt{0.79}$.

\subsection{Coherent plus squeezed vacuum input}
\label{subsec:Fisher_coh_sqz_vac}
In this scenario we have the input state
\begin{equation}
\label{eq:psi_in_coherent_plus_squeezed_vacuum}
\vert\psi_{in}\rangle=\vert\alpha_1\xi_0\rangle=\hat{D}_1\left(\alpha\right)\hat{S}_0\left(\xi\right)\vert0\rangle.
\end{equation}
The squeezed vacuum is obtained by applying the unitary operator \cite{GerryKnight,Yue76} 
\begin{equation}
\label{eq:Squeezing_operator}
\hat{S}_m\left(\chi\right)=e^{[\chi^*\hat{a}_m^2-\chi(\hat{a}_m^\dagger)^2]/2}
\end{equation}
to a mode $m$ previously found in the vacuum state and we denote $\chi=se^{i\vartheta}$. Usually $s\in\mathbb{R}^{+}$ is called the squeezing factor and ${\vartheta}$ denotes the phase of the squeezed state. For the input state from equation  \eqref{eq:psi_in_coherent_plus_squeezed_vacuum} we employed a squeezing with $\xi=re^{i\theta}$  applied to the input port $0$.
The first Fisher matrix element is $\mathcal{F}_{ss}=\vert\alpha\vert^2+\sinh^22r/2$. From equation \eqref{eq:F_dd_FINAL_FORM_GENERAL_compact} we have
\begin{eqnarray}
\label{eq:F_dd_coh_sqz_vac_nonbalanced_symm_BS}
\mathcal{F}_{dd}
=\left(\vert{T}\vert^2-\vert{R}\vert^2\right)^2\left(\vert\alpha\vert^2+\frac{\sinh^22r}{2}\right)
\nonumber\\
+4\vert{TR}\vert^2\left(\sinh^2r+\Upsilon^+\left(\alpha,\xi\right)\right)
\end{eqnarray}
where the $\Upsilon^+$ function is defined in equation \eqref{eq:function:Upsilon_Plus_Minus_Definition}. The last Fisher matrix element yields
\begin{eqnarray}
\label{eq:F_sd_coh_sqz_vac_nonbalanced_symm_BS}
\mathcal{F}_{sd}=
\left(\vert{T}\vert^2-\vert{R}\vert^2\right)
\left(\frac{\sinh^22r}{2}
-\vert\alpha\vert^2
\right)
\end{eqnarray}
and using equation \eqref{eq:F_2p_definition} we get the two-parameter QFI,
\begin{eqnarray}
\label{eq:Fisher_2p_coh_sqz_vac}
\mathcal{F}^{(2p)}=
4\vert{TR}\vert^2\left(\sinh^2r+\Upsilon^+\left(\alpha,\xi\right)\right)
\nonumber\\
+2\frac{\left(1-4\vert{TR}\vert^2\right)
\sinh^22r\vert\alpha\vert^2
}
{\vert\alpha\vert^2+\frac{\sinh^22r}{2}}
.
\end{eqnarray}
As discussed in reference \cite{Pre19}, if 
\begin{equation}
\label{eq:coh_plus_sqz_vac_T_opt_existence_condition}
\sinh^2r+\Upsilon^+\left(\alpha,\xi\right)
-\frac{2\sinh^22r\vert\alpha\vert^2}{\vert\alpha\vert^2+\frac{\sinh^22r}{2}}>0
\end{equation}
then $\mathcal{F}^{(2p)}$ is maximized in the balanced case and we get
\begin{eqnarray}
\label{eq:Fisher_2p_coh_sqz_vac_BALANCED}
\mathcal{F}^{(2p)}=\sinh^2r+\Upsilon^+\left(\alpha,\xi\right).
\end{eqnarray}
Imposing the optimum input PMC, namely,
\begin{equation}
\label{eq:PMC_coh_sqz_vac}
2\theta_\alpha-\theta=0
\end{equation}
implies $\Upsilon^+\left(\alpha,\xi\right)=\vert\alpha\vert^2e^{2r}$ and it maximizes the QFI to the well-known result $\mathcal{F}_{max}^{(2p)}=\sinh^2r+\vert\alpha\vert^2e^{2r}$ \cite{Pez07,Jar12,Dem15,Gar17}.

For the asymmetric single phase shift scenario from from Fig.~\ref{fig:MZI_Fisher_single_phase}, the single-parameter QFI \eqref{eq:Fisher_information_single_i} becomes
\begin{eqnarray}
\label{eq:F_i_coh_sqz_vac}
\mathcal{F}^{(i)}=
4\vert{T}\vert^4\vert\alpha\vert^2
+2\vert{R}\vert^4\sinh^22r
\nonumber\\
+4\vert{TR}\vert^2\left(\sinh^2r+\Upsilon^+\left(\alpha,\xi\right)\right).
\end{eqnarray}
$\mathcal{F}^{(i)}$ is maximized by an optimum transmission coefficient (see discussion in Appendix \ref{sec:app:Fisher_coh_plus_sqz_vac})
\begin{equation}
\label{eq:T_opt_coh_sqz_vac}
T_{opt}^{(i)}=\sqrt{\frac{\Upsilon^+\left(\alpha,\xi\right)-\sinh^2r(1+2\cosh2r)}
{2\left(\Upsilon^+\left(\alpha,\xi\right)-\vert\alpha\vert^2-\sinh^2r\cosh2r\right)}}
\end{equation}
The maximum single-parameter QFI, $\mathcal{F}^{(i)}_{max}$, can then be obtained by replacing $T_{opt}^{(i)}$ into equation \eqref{eq:F_i_coh_sqz_vac} and the result is given in equation \eqref{eq:F_i_coh_sqz_vac_MAX}.

\begin{figure}
\centering
\includegraphics[scale=0.45]{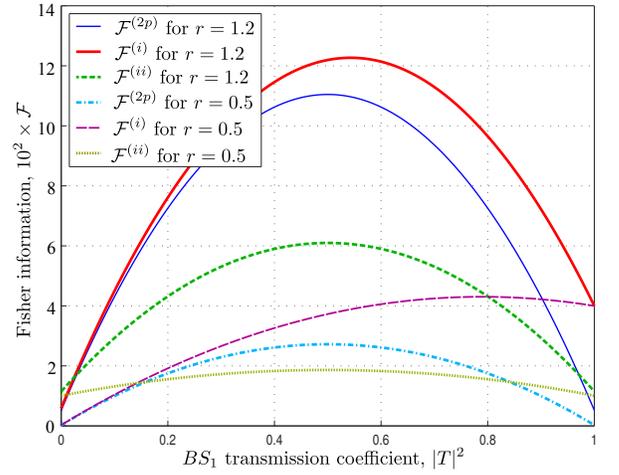}
\caption{The three considered QFIs versus the transmission coefficient of $BS_1$ for a coherent plus squeezed vacuum input state. As the squeezing factor $r$ increases, the quantum advantage of this state becomes obvious. Parameters used: $\vert\alpha\vert=10$ and the PMC $2\theta_\alpha-\theta=0$.}
\label{fig:Fisher_2p_i_ii_coh_sqz_vac}
\end{figure}

In the symmetrical case from Fig.~\ref{fig:MZi_Fisher_single_param_sym_varphi_over_2}, equation \eqref{eq:Fisher_information_single_ii} yields for the single parameter QFI
\begin{eqnarray}
\label{eq:F_ii_coh_sqz_vac}
\mathcal{F}^{(ii)}=\left(1-2\vert{TR}\vert^2\right)\left(\vert\alpha\vert^2+\frac{\sinh^22r}{2}\right)
\nonumber\\
+2\vert{TR}\vert^2\left(\sinh^2r+\Upsilon^+\left(\alpha,\xi\right)\right).
\end{eqnarray}
If the condition
\begin{equation}
\Upsilon^+\left(\alpha,\xi\right)-\vert\alpha\vert^2-\sinh^2r\cosh2r>0
\end{equation}
is satisfied, then $\mathcal{F}^{(ii)}$ is maximized in the balanced case yielding
\begin{eqnarray}
\label{eq:F_ii_coh_sqz_vac_MAX}
\mathcal{F}^{(ii)}_{max}=\frac{\vert\alpha\vert^2+\frac{\sinh^22r}{2}
+\sinh^2r+\Upsilon^+\left(\alpha,\xi\right)}{2}
\end{eqnarray}
The three considered QFIs are plotted in Fig.~\ref{fig:Fisher_2p_i_ii_coh_sqz_vac} versus the transmission coefficient of $BS_1$ for two squeezing factors. One notes the poor performance of $\mathcal{F}^{(ii)}$, even with respect to the two-parameter QFI. Both $\mathcal{F}^{(2p)}$ and $\mathcal{F}^{(ii)}$  yield their maximum value in the balanced scenario while $\mathcal{F}^{(i)}$ peaks at $T^{(i)}_{opt}$ given by equation \eqref{eq:T_opt_coh_sqz_vac}.

The ``quantum advantage'' becomes quite obvious if we compare Figs.~\ref{fig:Fisher_2p_i_ii_double_coh} and \ref{fig:Fisher_2p_i_ii_coh_sqz_vac}. Although $\vert\beta\vert^2>\sinh^2r$, the coherent plus squeezed vacuum completely outperforms the double coherent input in terms of maximum QFI.

If the condition $\vert\alpha\vert^2\gg\sinh^2r$ is satisfied and the optimum PMC \eqref{eq:PMC_coh_sqz_vac} employed, equation \eqref{eq:T_opt_coh_sqz_vac} approximates to
$T_{opt}^{(i)}\approx{e^{r}}/\sqrt{{2\left(e^{2r}-1\right)}}$
(valid under the constraint $T_{opt}^{(i)}\leq1$) implying the maximum single-parameter QFI,
\begin{equation}
\mathcal{F}^{(i)}_{max}\approx\frac{e^{4r}}{e^{2r}-1}|\alpha|^2.
\end{equation}
For small squeezing factors there is an advantage in employing the single parameter QFI (see Fig.~\ref{fig:Fisher_2p_i_ii_coh_sqz_vac}, dashed lines), a fact also seen from the fact that ${\mathcal{F}^{(i)}_{max}}/{\mathcal{F}^{(2p)}_{max}}\approx{e^{2r}}/{(e^{2r}-1)}\approx1.6$ for $r=0.5$.

For high squeezing factors we have $e^{2r}\gg1$, implying $T^{(i)}_{opt}\approx1/\sqrt{2}$. We thus have $\mathcal{F}^{(i)}_{max}\approx\mathcal{F}^{(2p)}_{max}\approx e^{2r}|\alpha|^2$. In other words, there is only a marginal advantage of having available an external phase reference in the high-intensity and strong squeezing regime for a coherent plus squeezed vacuum input.

\begin{figure}
		\includegraphics[scale=0.45]{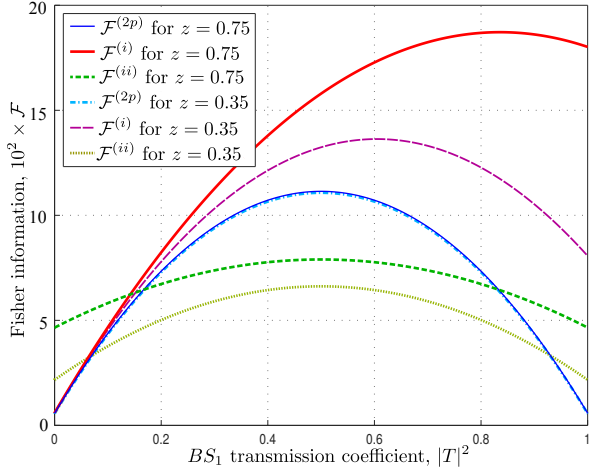}
	\caption{The three considered QFIs versus the transmission coefficient of $BS_1$ for a squeezed-coherent plus squeezed vacuum input state. The enhancement brought by the second squeezer is obvious for $\mathcal{F}^{(i)}$, however insignificant for $\mathcal{F}^{(2p)}$. Parameters used: $|\alpha|=10$, $r=1.2$, $2\theta_\alpha-\theta=0$, $2\theta_\alpha-\phi=\pi$.}
	\label{fig:Fisher_2p_i_ii_sqz-coh_sqz_vac_r_varied_alpha10}
\end{figure}

\subsection{Squeezed-coherent plus squeezed vacuum input}
\label{subsec:Fisher_sqz-coh_sqz_vac}
In this scenario we have the input state
\begin{equation}
\label{eq:psi_in_sqz-coh_sqz_vac}
\vert\psi_{in}\rangle=\vert(\alpha\zeta)_1\xi_0\rangle=\hat{D}_1\left(\alpha\right)\hat{S}_1\left(\zeta\right)\hat{S}_0\left(\xi\right)\vert0\rangle
\end{equation}
where we applied the squeezing operator \eqref{eq:Squeezing_operator} to port $1$ with the parameters $\zeta=ze^{i\phi}$. The first Fisher matrix element is found to be
\begin{eqnarray}
\label{eq:F_ss_sqz_coh_plus_sqz_vac_symm_BS_non_balanced}
\mathcal{F}_{ss}
=\frac{\sinh^22r}{2}
+\frac{\sinh^22z}{2}
+\Upsilon^-\left(\alpha,\zeta\right)
\end{eqnarray}
where the function $\Upsilon^-$ is defined by equation \eqref{eq:function:Upsilon_Plus_Minus_Definition}.
The other two Fisher matrix elements are detailed in Appendix
\ref{sec:app:Fisher_sqz_coh_plus_sqz_vac}. From these Fisher matrix elements we can compute the three considered QFIs, results also given in Appendix \ref{sec:app:Fisher_sqz_coh_plus_sqz_vac}.

The two-parameter QFI \eqref{eq:F2p_sqz-coh_sqz_vac_symm_BS_non_balanced} reduces to the simple expression $\mathcal{F}^{(2p)}=\vert\alpha\vert^2e^{2r}+\sinh^2(r+z)$ in the balanced case with the optimal input PMCs \cite{Pre19,Ata19}:
\begin{equation}
\label{eq:PMC_sqz-coh_plus_sqz-vac}
\left\{
\begin{array}{l}
2\theta_\alpha-\theta=0\\
\theta-\phi=\pm\pi.
\end{array}
\right.
\end{equation}
Similar to the coherent plus squeezed vacuum case, we can derive from equation  \eqref{eq:F_i_sqz-coh_plus_SQZ_VAC} an optimal transmission coefficient $T^{(i)}_{opt}$ that maximizes $\mathcal{F}^{(i)}$ (see Appendix \ref{sec:app:Fisher_sqz_coh_plus_sqz_vac}). The QFI describing the symmetrical $\pm\varphi/2$ is given in equation \eqref{eq:F_ii_sqz-coh_plus_SQZ_VAC} and if the condition \eqref{eq:Sqz-coh_plus_SQZ_VAC_CONDITION_F_ii_max_bal} is satisfied, it maximizes in the balanced case.

In Fig.~\ref{fig:Fisher_2p_i_ii_sqz-coh_sqz_vac_r_varied_alpha10} we plot the three QFIs versus the $BS_1$ transmission coefficient $|T|^2$. We considered two rather small squeezing factors ($z=0.35$ and, respectively, $z=0.75$). For the given parameters from Fig.~\ref{fig:Fisher_2p_i_ii_sqz-coh_sqz_vac_r_varied_alpha10}, we find $T^{(i)}_{opt}\approx \sqrt{0.6}$ (for $z=0.35$) and $T^{(i)}_{opt} \approx\sqrt{0.83}$ (for $z=0.75$).

In spite of being small, the squeezing from input port $1$ significantly enhances $\mathcal{F}^{(i)}$ around $T^{(i)}_{opt}$. These conclusions also hold in the experimentally interesting high-intensity regime $\vert\alpha\vert^2\gg\{\sinh^2r,\:\sinh^2z\}$, where we can approximate
\begin{equation}
\label{eq:T_opt_i_sqzcoh_sqzvac_high_alpha}
T^{(i)}_{opt}\approx
\sqrt{\frac{1}{2\vert1-e^{2(z-r)}\vert}}
\end{equation}
and this expression is meaningful as a transmission factor while $T^{(i)}_{opt}\leq1$ (see discussion in Appendix \ref{sec:app:Fisher_sqz_coh_plus_sqz_vac}).

We thus conclude that in both the low-intensity and high-intensity regimes, the availability of an external phase reference for a squeezed-coherent plus squeezed vacuum input brings a clear advantage.

\section{Realistic detection schemes}
\label{sec:realistic_detection_schemes}
\begin{figure}
\includegraphics[scale=0.55]{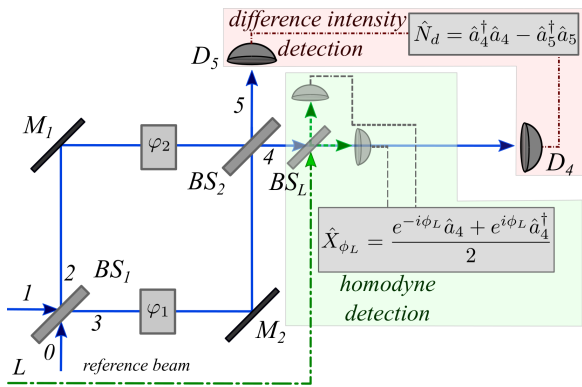}
\caption{The two realistic detection schemes considered here: the difference-intensity detection with its associated operator $\hat{N}_d$ and the balanced homodyne detection with its associated operator $\hat{X}_L$.}
\label{fig:Figure1_MZI_2D_differential_homodyne_v01}
\end{figure}

We close now the Mach-Zehnder interferometer (MZI) with $BS_2$ (characterized by the transmission/reflection coefficients $T'/R'$) and discuss the performance of two realistic detection schemes, namely the difference intensity  and the balanced homodyne detection techniques (see Fig.~\ref{fig:Figure1_MZI_2D_differential_homodyne_v01}). We consider the most general case (paralleling the two-parameter Fisher estimation from Section \ref{sec:Fisher_information_2param}) and assume two independent phase shifts $\varphi_1$ (in the lower arm of the interferometer) and $\varphi_2$ (in the upper one). Thus we can easily set $\varphi_2=0$ and we have the scenario from Section \ref{sec:Fisher_information_1param_asymetric} or we can set $\varphi_1=-\varphi_2=\varphi/2$ and find ourselves in the case described in Section \ref{sec:Fisher_information_1param_symetric_varphi_over2}.

\subsection{Difference-intensity detection}
\label{subsec:difference_detection_scheme}
A good example of a realistic detection scheme sensitive only to the difference phase shift ($\varphi_1-\varphi_2$)  is the difference-intensity detection scheme \cite{Gar17,API18,Ata19} (see Fig.~\ref{fig:Figure1_MZI_2D_differential_homodyne_v01}). The observable conveying information about the phase shift is
\begin{equation}
\label{eq:N_d_operator_DEFINITION}
\hat{N}_d=\hat{a}_4^\dagger\hat{a}_4-\hat{a}_5^\dagger\hat{a}_5
\end{equation}
and the final expression with respect to the input field operators is given in equation \eqref{eq:Nd_both_BS_unbalanced}.  The phase sensitivity is defined as usual,
\begin{equation}
\label{eq:Delta_varphi_diff_DEFINITION}
\Delta\varphi_{df}=\frac{\sqrt{\Delta^2\hat{N}_d}}{\Big\vert\frac{\partial\langle\hat{N}_d\left(\varphi\right)\rangle}{\partial\varphi}\Big\vert}
\end{equation}
where ${\partial\langle\hat{N}_d\left(\varphi\right)\rangle}/{\partial\varphi}$ is given in equation \eqref{eq:Nd_avg_del_varphi_both_BS_unbalanced} and the variance $\Delta^2\hat{N}_d$ is given in equation \eqref{eq:Variance_Nd_both_BS_unbalanced}.

\subsection{Balanced homodyne detection}
\label{subsec:homodyne_detection_}
If we assume a balanced homodyne detection scheme at the output port $4$ (see Fig.~\ref{fig:Figure1_MZI_2D_differential_homodyne_v01}), the relevant operator modeling this detection is given by
\begin{equation}
\hat{X}_{\phi_L}=\frac{e^{-i\phi_L}\hat{a}_4+e^{i\phi_L}\hat{a}_4^\dagger}{2}
\end{equation}
where $\phi_L$ (assumed fixed and controllable with respect to $\theta_\alpha$) is the phase of the local coherent source $\ket{\gamma}$ where $\gamma=\vert\gamma\vert e^{i\phi_L}$. The final expression for $\hat{X}_{\phi_L}$ with regard to the input field operators is given in Appendix \ref{sec:app:homodyne}. We define the phase sensitivity of a balanced homodyne detector as
\begin{equation}
\label{eq:Delta_varphi_hom_DEFINITION}
\Delta\varphi_{hom}=\frac{\sqrt{\Delta^2\hat{X}_{\phi_L}}}{\Big\vert\frac{\partial\langle\hat{X}_{\phi_L}\rangle}{\partial\varphi}\Big\vert}
\end{equation}
If we consider the scenario from Fig.~\ref{fig:MZI_Fisher_single_phase}, we have
\begin{eqnarray}
\label{eq:del_X_phi_L_avg__del_varphi_homodyne_non_bal_one_phase}
\bigg\vert\frac{\langle\hat{X}_{\phi_L}\rangle}{\partial\varphi}\bigg\vert
=\Big\vert\Re\left\{e^{-i(\phi_L+\varphi)}\left(R\langle{\hat{a}_0}\rangle
+T\langle{\hat{a}_1}\rangle\right)\right\}\Big\vert|R'|
\end{eqnarray}
and for the symmetric $\pm\varphi/2$ scenario from Fig.~\ref{fig:MZi_Fisher_single_param_sym_varphi_over_2} we get
\begin{eqnarray}
\label{eq:del_X_phi_L_avg__del_varphi_homodyne_non_bal_two_phases}
\bigg\vert\frac{\partial\langle\hat{X}_{\phi_L}\rangle}{\partial\varphi}\bigg\vert
=\frac{1}{2}\Re\left\{ie^{-i\phi_L}\left(TT'e^{i\varphi/2}-RR'e^{-i\varphi/2}\right)\langle{\hat{a}_0}\rangle
\right.
\nonumber\\
\left.
+ie^{-i\phi_L}\left(RT'e^{i\varphi/2}-TR'e^{-i\varphi/2}\right)\langle{\hat{a}_1}\rangle\right\}
.
\quad\:\:
\end{eqnarray}
The final expression for the variance $\Delta^2\hat{X}_{\phi_L}$ with respect to the input field operators is given in equation \eqref{eq:variance_X_phi_L_homodyne}.

\section{Phase sensitivity comparison with Gaussian input states}
\label{sec:phase_sensitivity_gaussian_states}
Paralleling the discussion from Section \ref{sec:Fisher_input_Gaussian_states}, we compare here the realistically achievable phase sensitivities for various input Gaussian states versus the QCRBs implied by the QFIs discussed before. We consider both detection schemes presented in Section \ref{sec:realistic_detection_schemes}.

\subsection{Single coherent input}
\label{subsec:phase_sensitivity_gaussian_states_single_coh}
From the phase sensitivity formula \eqref{eq:Delta_varphi_diff_DEFINITION} and considering the input state \eqref{eq:psi_in_single_coh}, for a difference-intensity detection scheme we get
\begin{equation}
\label{eq:Delta_varphi_diff_single_coh}
\Delta\varphi_{df}=\frac{1}{4\vert TRT'R'\vert\vert\alpha\vert\vert\sin\varphi\vert}
\end{equation}
and comparing this result with the QCRB from equation \eqref{eq:Fisher_2p_single_coherent} we note that it can be attained only if $BS_2$ is balanced. Moreover, this detection scheme yields the same result for the scenarios from Figs.~\ref{fig:MZI_Fisher_single_phase} and \ref{fig:MZi_Fisher_single_param_sym_varphi_over_2}. The phase sensitivity $\Delta\varphi_{df}$ from equation  \eqref{eq:Delta_varphi_diff_single_coh} is further optimized if $BS_1$ is balanced, too, yielding the well-known result \cite{Dem15,API18}
\begin{equation}
\label{eq:Delta_varphi_diff_single_coh_BEST}
\Delta\varphi_{df}=\frac{1}{\vert\alpha\vert\vert\sin\varphi\vert}.
\end{equation}
We note that for $|T|\to1$ (or $|T|\to0$) the phase sensitivity degrades, a behavior expected from the vanishing of the two-parameter QFI from equation \eqref{eq:F_2p_T_equal_0_F_sd_not_zero}.

For a balanced homodyne detection scheme we obtain the variance $\Delta^2\hat{X}_{\phi_L}=1/4$.
In the setup from Fig.~\ref{fig:MZI_Fisher_single_phase}, from equation \eqref{eq:del_X_phi_L_avg__del_varphi_homodyne_non_bal_one_phase} we have
\begin{eqnarray}
\label{eq:del_X_phi_L_avg__del_varphi_homodyne_non_bal_one_phase_single_coh}
\bigg\vert\frac{\langle\hat{X}_{\phi_L}\rangle}{\partial\varphi}\bigg\vert
=|TR'|\vert\alpha\vert\vert\cos(\phi_L+\varphi-\theta_\alpha)\vert
\end{eqnarray}
and imposing $\phi_L=\theta_\alpha$ we end up with a phase sensitivity
\begin{equation}
\label{eq:Delta_varphi_homodyne_i_single_coh}
\Delta\varphi^{(i)}_{hom}=\frac{1}{2|TR'|\vert\alpha\vert\vert\cos\varphi\vert}.
\end{equation}
Contrary to $\Delta\varphi_{df}$ from equation \eqref{eq:Delta_varphi_diff_single_coh}, an unbalanced interferometer with $|T|\to1$ and $|T'|\to0$ actually takes us to $\Delta\varphi^{(i)}_{QCRB}$ from equation \eqref{eq:Delta_varphi_QCRB_i_DEFINTION} if we select the optimal working point where $\vert\cos\varphi\vert=1$. In the balanced case, the best phase sensitivity $\Delta\varphi^{(i)}_{hom}={1}/{\vert\alpha\vert}$  is indeed limited by $\Delta\varphi_{QCRB}^{(2p)}$ and this explains why previous papers \cite{Gar17,Ata19} did not report sensitivities beyond the QCRB from equation \eqref{eq:Delta_varphi_QCRB_2p_DEFINTION}.

\begin{figure}
		\includegraphics[scale=0.45]{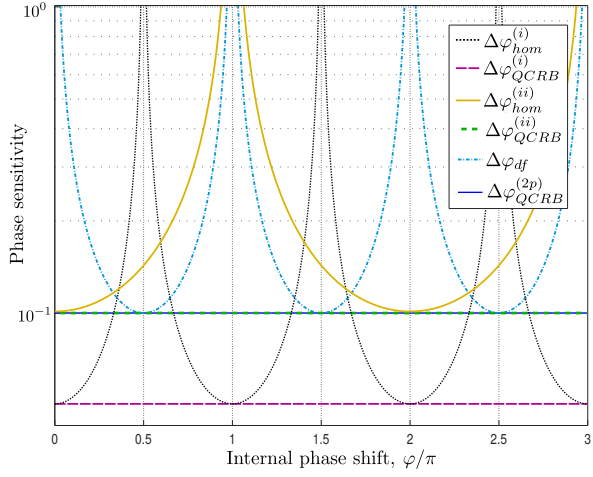}
	\caption{Phase sensitivity for a single coherent input. The balanced homodyne detection scheme approaches the QCRBs corresponding to $\mathcal{F}^{(i)}$ and, respectively $\mathcal{F}^{(ii)}$. The performance of the difference intensity detection scheme is limited by the QCRB corresponding to the two-parameter QFI. Parameters used: $\vert\alpha\vert=10$ and $\phi_L=\theta_\alpha$.}
	\label{fig:Delta_varphi_single_coh_alpha10_Nd_bal_hom_T_99}
\end{figure}

In the symmetrical scenario (see Fig.~\ref{fig:MZi_Fisher_single_param_sym_varphi_over_2}), from equation \eqref{eq:del_X_phi_L_avg__del_varphi_homodyne_non_bal_two_phases} we get
\begin{eqnarray}
\label{eq:del_X_phi_L_avg__del_varphi_homodyne_non_bal_two_phases_single_coh}
\bigg\vert\frac{\partial\langle\hat{X}_{\phi_L}\rangle}{\partial\varphi}\bigg\vert
=\frac{\vert\alpha\vert}{2}\big\vert|TR'|\cos\left(\phi_L+\varphi/2-\theta_\alpha\right)
\nonumber\\
-|T'R|\cos(\phi_L-\varphi/2-\theta_\alpha)\big\vert
\end{eqnarray}
and with the condition $\phi_L-\theta_\alpha=0$ we find the phase sensitivity
\begin{equation}
\label{eq:Delta_varphi_homodyne_ii_single_coh}
\Delta\varphi^{(ii)}_{hom}=\frac{1}{\big\vert|TR'|
-|RT'|\big\vert\vert\alpha\vert\big|\cos\left(\frac{\varphi}{2}\right)\big|}
\end{equation}
It is obvious that $\max\big\vert|TR'|
-|RT'|\big\vert\leq1$ and this limit is saturated for $|T|\to1$ and $|T'|\to0$ or for $|T|\to0$ and $|T'|\to1$. Assuming the first case we have $|TR'|\approx1$ and $|T'R|\approx0$ thus we get the best phase sensitivity at the optimum angle ($\varphi_{opt}=2k\pi$ with $k\in\mathbb{Z}$)
\begin{equation}
\label{eq:Delta_varphi_homodyne_ii_single_coh_BEST}
\Delta\tilde{\varphi}^{(ii)}_{hom}\approx\frac{1}{\vert\alpha\vert}
\end{equation}
This phase sensitivity is indeed limited by the QCRB $\Delta{\varphi}^{(ii)}_{QCRB}$ from equation \eqref{eq:Delta_varphi_QCRB_ii_DEFINTION}.

In Fig.~\ref{fig:Delta_varphi_single_coh_alpha10_Nd_bal_hom_T_99} we depict the performance of both detection schemes versus the three QCRBs discussed before. We plot the difference-intensity detection scheme at its best performance (implying both BS balanced). For the balanced homodyne detection scheme we consider the transmission coefficients  $T=0.99$  (for $BS_1$) and $T'=0.01$ (for $BS_2$). As see from Fig.~\ref{fig:Delta_varphi_single_coh_alpha10_Nd_bal_hom_T_99}, all three QCRBs have a physical meaning and with the appropriate setup can actually be attained.

\begin{figure}
		\includegraphics[scale=0.45]{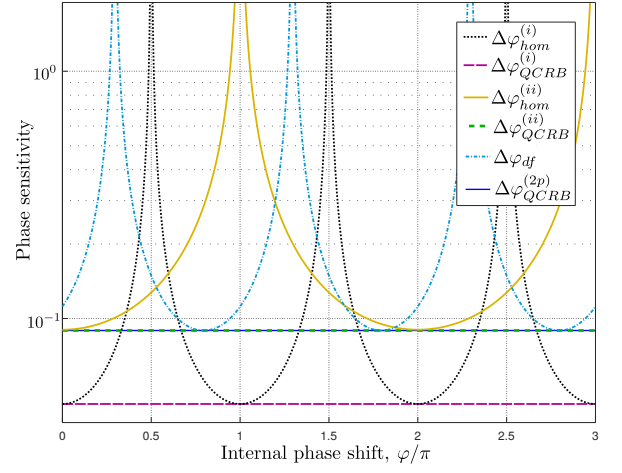}
	\caption{Phase sensitivity for a dual coherent input. The homodyne detection approaches the QCRBs corresponding to the single-parameter QFI. Parameters used: $\vert\alpha\vert=10$, $\vert\beta\vert=5$. The difference-intensity detection scheme reaches the QCRB corresponding to the two-parameter QFI. Please note the different optimal working points for the considered detection schemes.}
	\label{fig:Delta_varphi_dual_coh_alpha10_beta5_Nd_bal_hom_T_opt}
\end{figure}

\subsection{Double coherent input}
\label{subsec:phase_sensitivity_gaussian_states_dual_coh}
With a double coherent input state \eqref{eq:psi_in_dual_coh} more degrees of freedom become available in order to outline the physical meaning of each of the three QCRB.

In the case of a difference-intensity detection scheme we have the phase sensitivity (see Appendix \ref{sec:app:phase_sensitivity_dual_coh})
\begin{equation}
\label{sec:Delta_varphi_df_dual_coh}
\Delta\varphi_{df}
=\frac{\sqrt{\vert\alpha\vert^2+\vert\beta\vert^2}}{4\big\vert|TR|\sin\varphi(\beta\vert^2-\vert\alpha\vert^2)
+\vert\alpha\beta\vert\cos\varphi\big\vert\vert T'R'\vert}
\end{equation}
and we assumed here the optimal PMC $\Delta\theta=0$. The phase sensitivity is further optimized for both BS balanced yielding its best performance at the working point 
\begin{equation}
\label{sec:Delta_varphi_df_dual_coh_OPTIMAL}
\Delta\tilde{\varphi}_{df}=\frac{1}{\sqrt{\vert\alpha\vert^2+\vert\beta\vert^2}}
\end{equation}
where the optimum working point $\varphi_{opt}$ is given by equation \eqref{eq:varphi_OPT_dual_coh} and we conclude that we can attain the $\Delta\varphi_{QCRB}^{(2p)}$ from equation \eqref{eq:QCRB_F_2p_dual_coh}. We wish to point out that in general, $\varphi_{opt}\neq{k}\pi/2$ with $k\in\mathbb{Z}$.

We emphasize now an interesting point about the dual coherent input state, already mentioned in Section \ref{subsec:Fisher_input_Gaussian_states_dual_coh}, namely that if we change the input PMC from $\Delta\theta=0$ to $\Delta\theta=\pi/2$, $\mathcal{F}^{(2p)}$ is decreased, however $\mathcal{F}^{(i)}$ increases. While the decrease of $\mathcal{F}^{(2p)}$ is less surprising and already discussed in the literature \cite{API18,Pre19}, the increase of $\mathcal{F}^{(i)}$ is somehow surprising and the attainability of its corresponding QCRB may raise some doubts.

For a balanced homodyne detection scheme we get for the variance $\Delta^2\hat{X}_{\phi_L}=1/4$. In the case of an asymmetric phase shift (see Fig.~\ref{fig:MZI_Fisher_single_phase}), from equation \eqref{eq:del_X_phi_L_avg__del_varphi_homodyne_non_bal_one_phase} we get
\begin{eqnarray}
\bigg\vert\frac{\langle\hat{X}_{\phi_L}\rangle}{\partial\varphi}\bigg\vert
=\big\vert |T|\cos\varphi\vert\alpha\vert 
+|R|\sin(\Delta\theta+\varphi)\vert\beta\vert 
\big\vert|R'|
\end{eqnarray}
and we assumed $\phi_L=\theta_\alpha$.  If we further assume $\Delta\theta=\pi/2$ and impose the optimal transmission factor $T^{(i)}_{opt}$ from equation \eqref{eq:T_opt_Delta_theta_pi_over_2_dual_coh_F_i}, we get the phase sensitivity
\begin{equation}
\Delta\varphi_{hom}^{(i)}=\frac{1}{2|R'|\sqrt{\vert\alpha\vert^2+\vert\beta\vert^2}|\cos\varphi|}
\end{equation}
and for at the optimum working point $\varphi_{opt}=(2k+1)\pi$ ($k\in\mathbb{Z}$) and $|R'|\to1$ we have indeed $\Delta\varphi_{hom}^{(i)}\to\Delta\varphi_{QCRB}^{(i)}$.

In the case of a symmetric phase shift (see Fig.~\ref{fig:MZi_Fisher_single_param_sym_varphi_over_2}), from equation \eqref{eq:del_X_phi_L_avg__del_varphi_homodyne_non_bal_two_phases} and imposing $\phi_L=\theta_\alpha$, the optimal transmission factor from equation \eqref{eq:T_opt_Delta_theta_pi_over_2_dual_coh_F_i} and $\Delta\theta=\pi/2$ we get
\begin{equation}
\Delta\varphi_{hom}^{(ii)}=\frac{1}{|R'|\sqrt{\vert\alpha\vert^2+\vert\beta\vert^2}\big|\cos\frac{\varphi}{2}\big|}
\end{equation}
and we assumed $|R'|\gg|T'|$. Imposing the optimum working point $\varphi_{opt}=2k\pi$ ($k\in\mathbb{Z}$) and $|R'|\to1$ we have $\Delta\varphi_{hom}^{(ii)}\to\Delta\varphi_{QCRB}^{(ii)}$.

In Fig.~\ref{fig:Delta_varphi_dual_coh_alpha10_beta5_Nd_bal_hom_T_opt} we plot the three phase sensitivities against their corresponding QCRBs. Similar to Section \ref{subsec:phase_sensitivity_gaussian_states_single_coh}, the difference-intensity detection scheme is considered with both BS balanced. For the balanced homodyne detection we consider $T$ given by equation \eqref{eq:T_opt_Delta_theta_pi_over_2_dual_coh_F_i} and for the second BS we took $T'=0.01$. One notes that each detection scheme approaches its corresponding QCRB.

\subsection{Coherent plus squeezed vacuum input}
\label{subsec:phase_sensitivity_gaussian_states_coh_sqz_vac}
In the following two sections we only consider the phase sensitivities $\Delta\varphi_{hom}^{(i)}$ and $\Delta\varphi_{hom}^{(2p)}$. With the input state given by equation \eqref{eq:psi_in_coherent_plus_squeezed_vacuum} we find 
\begin{eqnarray}
\bigg\vert\frac{\partial\langle\hat{N}_d\rangle}{\partial\varphi}\bigg\vert=
4\vert TRT'R'\vert \vert\sin\varphi\vert \big\vert\vert\alpha\vert^2-\sinh^2r\big\vert
\end{eqnarray}
One notes that the balanced case (for both $BS$) maximizes this term. For the variance $\Delta^2\hat{N}_d$,  we obtain the result given in equation \eqref{eq:Variance_Nd_both_BS_unbalanced_coh_sqz_vac}.

In the case of a balanced homodyne detection, equation \eqref{eq:del_X_phi_L_avg__del_varphi_homodyne_non_bal_one_phase_single_coh} remains valid. The variance is given by equation \eqref{eq:variance_X_phi_L_homodyne_symbolic} and combining these results takes us to the phase sensitivity from equation \eqref{eq:Delta_varphi_hom_coh_sqz_vac_symbolic}. Further simplifications are obtained by assuming $\phi_L=\theta_\alpha$ and the PMC \eqref{eq:PMC_coh_sqz_vac} satisfied, yielding the phase sensitivity $\Delta{\varphi}_{hom}^{(i)}$ from equation
\eqref{eq:Delta_varphi_hom_coh_sqz_vac_PMC_optimal}.

Imposing the optimum working point $\varphi_{opt}=\pi$ takes us to the best achievable phase sensitivity,
\begin{equation}
\label{eq:Delta_varphi_hom_coh_sqz_vac_BEST}
\Delta\tilde{\varphi}_{hom}^{(i)}=\frac{\sqrt{1
-(|TT'|+|RR'|)^2(1-e^{-2r}) 	
}}
{2|TR'|\vert\alpha\vert}.
\end{equation}
We notice that we get the well-known result $\Delta\tilde{\varphi}_{hom}^{(i)}=e^{-r}/\vert\alpha\vert$ \cite{Gar17,Ata19} by imposing both BS balanced. However, this is not the optimum setup. The best phase sensitivity is obtained by imposing the transmission coefficient ${T}^{(i)}_{opt}$ from equation \eqref{eq:T_opt_coh_sqz_vac} to $BS_1$ and
\begin{equation}
\label{eq:Tprime_opt_coh_sqz_vac}
{T'}^{(i)}_{opt}
=\frac{{T}^{(i)}_{opt}\sqrt{1-\left({T}^{(i)}_{opt}\right)^2}(1-e^{-2r})}
{\sqrt{1-\left({T}^{(i)}_{opt}\right)^2\left(1-e^{-4r}\right)}}
\end{equation}
to $BS_2$.

\begin{figure}
	\includegraphics[scale=0.45]{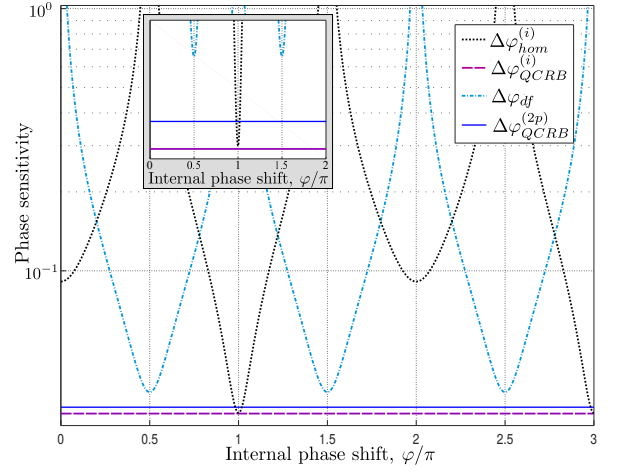}
	\caption{Phase sensitivity for a coherent plus squeezed vacuum input. Parameters used: $\vert\alpha\vert=10$, $r=1.2$ and PMC $2\theta_\alpha-\theta=0$. Inset: zoom around the peak sensitivity for the phase shift range $\varphi/\pi\in[0,2]$. While $\Delta\varphi_{df}$ is largely suboptimal, $\Delta\varphi^{(i)}_{hom}$ is much closer to optimality.}
	\label{fig:Delta_phi_coh_sqz-vac_alpha10_r1_2}
\end{figure}

In Fig.~\ref{fig:Delta_phi_coh_sqz-vac_alpha10_r1_2} we depict the phase sensitivities as well as the corresponding QCRBs versus the internal phase shift. For the difference-intensity detection scheme we considered its optimal setup, i. e. with both BS balanced. For the homodyne  detection scheme we considered the optimal transmission coefficients for the parameters used in Fig.~\ref{fig:Delta_phi_coh_sqz-vac_alpha10_r1_2}, namely $T_{opt}^{(i)}\approx\sqrt{0.53}$ and ${T'}^{(i)}_{opt}\approx\sqrt{0.44}$.

\begin{figure}
	\includegraphics[scale=0.45]{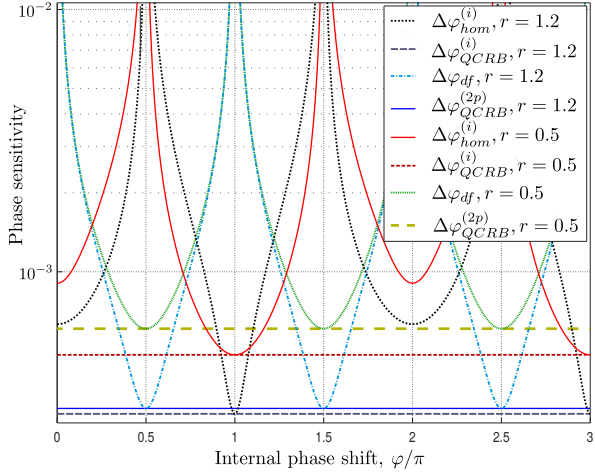}
	\caption{Phase sensitivity for coherent plus squeezed vacuum input in the high-$\alpha$ regime. Parameters used: $\vert\alpha\vert=10^3$,  $2\theta_\alpha-\theta=0$ and $\theta-\phi=\pi$.}
	\label{fig:Delta_phi_coh_sqz-vac_alpha1000_r0_5_1_2}
\end{figure}

The experimentally interesting high-$\alpha$ regime is depicted in Fig.~\ref{fig:Delta_phi_coh_sqz-vac_alpha1000_r0_5_1_2}. For small squeezing ($r=0.5$ in our case), $\Delta\varphi_{hom}^{(i)}$ approaches $\Delta\varphi_{QCRB}^{(i)}$ and shows noticeably better performance than the $\Delta\varphi_{QCRB}^{(2p)}$. However, for a higher squeezing factor ($r=1.2$ in our case) the two QFIs as well as the two detection schemes yield an almost similar performance (solid lines in Fig.~\ref{fig:Delta_phi_coh_sqz-vac_alpha1000_r0_5_1_2}).

We conclude that for a coherent plus squeezed vacuum input state there is a certain advantage of using an external phase reference in the high-$\alpha$ regime with the constraint of a low squeezing factor.

\subsection{Squeezed-coherent plus squeezed vacuum input}
\label{subsec:phase_sensitivity_gaussian_states_sqz-coh_sqz_vac}

For the input state from equation \eqref{eq:psi_in_sqz-coh_sqz_vac} and a difference-intensity detection scheme we find
\begin{eqnarray}
\label{eq:del_Nd_del_varphi_non_bal_coh_sqz_vac}
\bigg\vert\frac{\partial\langle\hat{N}_d\rangle}{\partial\varphi}\bigg\vert=
4\vert TRT'R'\vert \big\vert\vert\alpha\vert^2+\sinh^2z-\sinh^2r\big\vert
\vert\sin\varphi\vert
\quad\quad
\end{eqnarray}
The variance $\Delta^2\hat{N}_d$ can be obtained as before by applying the input state \eqref{eq:psi_in_sqz-coh_sqz_vac} to equation \eqref{eq:Variance_Nd_both_BS_unbalanced}. Throughout this section we consider the input PMCs \eqref{eq:PMC_sqz-coh_plus_sqz-vac} satisfied. The balanced case for both $BS_1$ and $BS_2$ maximizes equation \eqref{eq:del_Nd_del_varphi_non_bal_coh_sqz_vac}. The expression of $\Delta^2\hat{N}_d$ for the balanced case can be found in reference \cite{Ata19}.

For a balanced homodyne detection scheme, equation \eqref{eq:del_X_phi_L_avg__del_varphi_homodyne_non_bal_one_phase_single_coh} remains valid. For   $\phi_L=\theta_\alpha$ and PMCs \eqref{eq:PMC_sqz-coh_plus_sqz-vac} satisfied, the variance $\Delta^2\hat{X}_L$ is given in  equation \eqref{eq:Variance_X_sqz_coh_squeezed_vacuum_FINAL}. Combining these findings and imposing the optimum working point $\varphi_{opt}=\pi$ takes us to the optimal phase sensitivity from equation \eqref{eq:Delta_varphi_sqz_coh_squeezed_vacuum_varphi_OPTIMAL}.

\begin{figure}
		\includegraphics[scale=0.45]{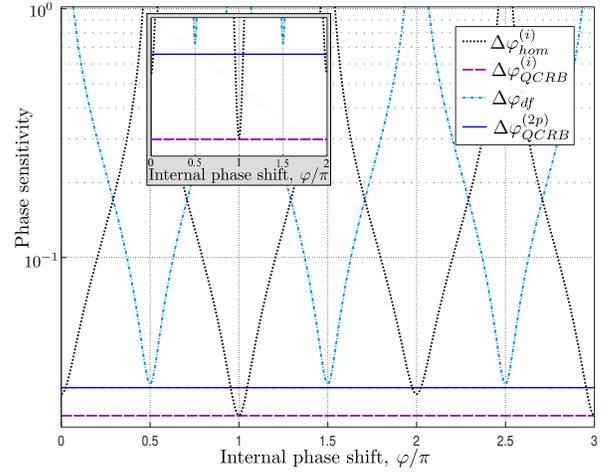}
	\caption{Phase sensitivity for a squeezed-coherent plus squeezed vacuum input versus the internal phase shift. Even with a modest squeezing factor in port $1$, the advantage of using a squeezed-coherent plus squeezed vacuum input state is obvious. Parameters used: $\vert\alpha\vert=10$, $r=1.2$, $z=0.75$ and PMCs $2\theta_\alpha-\theta=0$, $\theta-\phi=\pi$. Inset: zoom around the peak sensitivity for the phase shift range $\varphi/\pi\in[0,2]$. While $\Delta\varphi_{df}$ is still suboptimal, $\Delta\varphi^{(i)}_{hom}$ is very close to optimality.}
	\label{fig:Delta_phi_sqz-coh_sqz_vac_alpha_100_r15_z12_vs_theta}
\end{figure}

In Fig.~\ref{fig:Delta_phi_sqz-coh_sqz_vac_alpha_100_r15_z12_vs_theta} we plot the performance of both detectors versus the internal phase shift. For the difference intensity detection scheme we considered both BS balanced while in the case of the balanced homodyne detection we applied the optimal transmission factors $T^{(i)}_{opt}\approx\sqrt{0.62}$ for $BS_1$ and ${T'}^{(i)}_{opt}\approx\sqrt{0.17}$ for $BS_2$. We recall that $T_{opt}^{(i)}$ stems from optimizing the QFI $\mathcal{F}^{(i)}$, while ${T'}^{(i)}_{opt}$ was obtained by minimizing $\Delta\varphi_{hom}^{(i)}$ and the result is given by equation \eqref{eq:T_prime_opt_sqzcoh_plus_sqzvac}.

As already noted in Section \ref{subsec:Fisher_coh_sqz_vac}, for a coherent plus squeezed vacuum input, in the high-$\alpha$ regime with small squeezing, there is a certain advantage in using an external phase reference. However as the squeezing factor increases, we have $\mathcal{F}^{(i)}_{max}\approx\mathcal{F}^{(2p)}_{max}$. This fact changed here, irrespective of the squeezing factor from port $0$, there is a sizable increase in the performance for a squeezed-coherent plus squeezed vacuum input state.

\begin{figure}
	\includegraphics[scale=0.45]{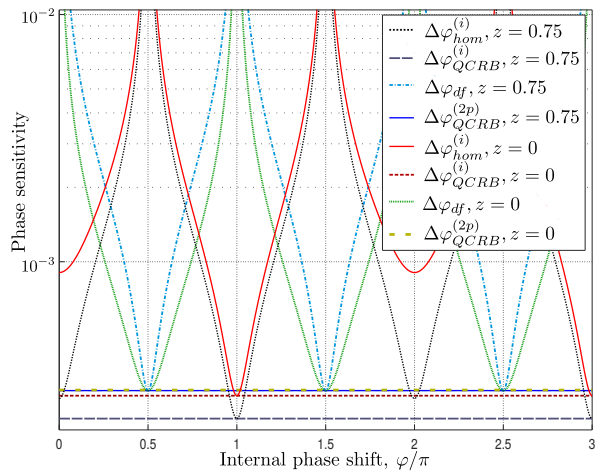}
	\caption{Phase sensitivity comparison between coherent plus squeezed vacuum and squeezed-coherent plus squeezed vacuum input in the high-$\alpha$ regime. Parameters used: $\vert\alpha\vert=10^3$, $r=1.2$, $2\theta_\alpha-\theta=0$ and $\theta-\phi=\pi$.}
	\label{fig:Delta_phi_coh_sqz-vac_and_sqzcoh_sqvac_alpha1000_r1_2_z075}
\end{figure}

In order to better outline this assertion, in Fig.~\ref{fig:Delta_phi_coh_sqz-vac_and_sqzcoh_sqvac_alpha1000_r1_2_z075} we plot on the same graphic the performance of coherent plus squeezed vacuum (i. e. $z=0$) and squeezed-coherent plus squeezed vacuum inputs in the high-$\alpha$ regime. 

Thus, we conclude that having access to an external phase reference for a squeezed-coherent plus squeezed vacuum input state brings a gain in the phase sensitivity, gain that does not fade away with the increase of the coherent amplitude, $\vert\alpha\vert$.

\section{Discussion}
\label{sec:discussion}
For a single coherent input state and a balanced homodyne detection scheme we obtained in Section \ref{subsec:phase_sensitivity_gaussian_states_single_coh} the ``unphysical'' limits $|T|\to1$ and $|T'|\to0$ in order to reach the bound $\Delta\varphi^{(i)}_{QCRB}=1/2|\alpha|$ implied by $\mathcal{F}^{(i)}$. However, by analyzing Fig.~\ref{fig:Figure1_MZI_2D_differential_homodyne_v01}, there is absolutely nothing unphysical about these limits. Indeed, we can write the input state (including the local oscillator) as 
\begin{equation}
\ket{\Psi}=\ket{\psi_{in}}\otimes\ket{\gamma}=\ket{e^{i\theta_\alpha}\vert\alpha\vert}\otimes\ket{e^{i\phi_L}\vert\gamma\vert}
\end{equation}
and we have an interferometer with one arm comprising the input port $1$, through $BS_1$ (total transmission), phase shift $\varphi_1$, $BS_2$ (total reflection) and to $BS_L$ while the other arm is simply the local oscillator fed into the homodyne's balanced beam splitter. Since the two input signals have a fixed phase relation, interference is to be expected. In the case of the dual coherent input from Section \ref{subsec:phase_sensitivity_gaussian_states_dual_coh}, $BS_1$ is no more in total transmission/reflection mode, its transmission coefficient $T^{(i)}_{opt}$ being given by equation \eqref{eq:T_opt_Delta_theta_pi_over_2_dual_coh_F_i}. When squeezing is added in one or both inputs, $\Delta\varphi^{(2p)}_{QCRB}$ can be outperformed and $\Delta\varphi^{(i)}_{QCRB}$ approached with both $BS_1$ and $BS_2$ having well defined values of their respective transmission coefficients ($|T|,|T'|\neq\{0,1\}$), as discussed in the previous sections.


In reference \cite{Tak17} it was claimed that: \emph{``First, if both arms of the MZI have different unknown phase shifts in the application and the input to one of the two ports is vacuum, then no matter what the input in the other port is, and no matter the detection scheme, one can never better the SNL in phase sensitivity. {\normalfont[\ldots]} This type of sensing includes gravitational
wave detection {\normalfont[\ldots]}''}. Indeed, treating this as a two-parameter problem and imposing the vacuum state for input port $0$, equation \eqref{eq:F_2p_definition} yields $\mathcal{F}^{(2p)}=4|TR|^2\langle\hat{n}_1\rangle$ and  since $4|TR|^2\leq1$ there is no room for sub-SNL performance. However, we would like to point out a small exception to this rule, namely if the two unknown phase shifts are correlated ($\varphi_1=\varphi/2$ and $\varphi_2=-\varphi/2$), then $\mathcal{F}^{(ii)}$ applies, not $\mathcal{F}^{(2p)}$. With the input port $0$ in the vacuum state, equation \eqref{eq:Fisher_information_sg_param_symmetrical_varphi_over_2} yields 
\begin{equation}
\mathcal{F}^{(ii)}=
(\vert{T}\vert^4+\vert{R}\vert^4)\Delta^2{\hat{n}_1}
+2\vert{TR}\vert^2\langle{\hat{n}_1}\rangle
\end{equation}
and we have a sub-SNL sensitivity if $\Delta^2{\hat{n}_1}>\langle{\hat{n}_1}\rangle$. We would also like to point out that gravitational waves with the $+$ polarization along any of the arms of the detector and arriving perpendicular to the plane of the interferometer yield highly (anti)correlated phase shifts \cite{Abbott_2009,LIGO2018}.


It has been argued that the external phase reference (the homodyne in our case) must be strong compared to the other sources (e. g. $\vert\gamma\vert^2\gg\{\vert\alpha\vert^2,\sinh^2r\}$ for a coherent plus squeezed vacuum input). Thus, one might object that this scheme is irrelevant since it requires even more resources. There are two arguments against this objection. Sometimes the sample that causes the phase shift inside the interferometer is delicate, as in the case of a microscope \cite{Tay13,Ono13}. Thus, the available power shone on the sample has to be drastically limited and any phase sensitivity enhancement via an external phase reference is more than welcome. Second, when the interferometer is at its optimum working point, the average photon number at output port $4$ (for the single-mode intensity and balanced homodyne detection schemes) is low for many input states \cite{API18}. Thus, the external phase reference can actually have a much lower amplitude than initially anticipated.


Although losses are outside the scope of this paper, we can schematically discuss the effects of non-ideal photon detectors  \cite{Kim99,Spa16,Ono10,Ata19} and/or internal losses \cite{Dor09,Dem09,Ono10}. Non-ideal photo-detection can be modeled by inserting a ficticious BS with a transmission factor $\eta$ ($\eta=1$ implies no losses) in front of a ideal photo-detector \cite{Kim99,Spa16,Ono10,Ata19}. In the case of coherent states, we have the scaling $\Delta\varphi_{lossy}=1/\sqrt{\eta}\Delta\varphi_{ideal}$ \cite{Ata19,Kim99,Spa16}. Thus, for modern, high-efficient photo-detectors the effect should be marginal. The impact is more severe in the case of a coherent plus squeezed vacuum input \cite{Ono10,Ata19,Spa16} and in the case of high losses the scaling approaches the SNL. For a squeezed-coherent plus squeezed vacuum input a similar pattern emerges \cite{Ata19}. However, workarounds have been shown to exist. Wu, Toda \& Hofmann \cite{Wu19} showed that by using photon-number-resolving detectors (PNRDs) in the dark port of an interferometer fed by a coherent plus squeezed vacuum input, up to a certain level of losses, the quantum Cram\'er-Rao bound can be attained. In the case of coherent light input, internal losses have the same effect as non-ideal photodetectors, while for a coherent plus squeezed vacuum input state they impact more the QFI terms that could have lead to a Heisenberg scaling \cite{Ono10}.

\section{Conclusions}
\label{sec:conclusions}
In this paper we reconsidered the single-parameter QFI versus the two-parameter one for an unbalanced MZI. We theoretically calculated the single parameter QFI both for a asymmetric and symmetric phase shifts scenarios as well as the two-parameter QFI. From these QFIs we can infer their corresponding quantum Cram\'er-Rao bounds, implying the best achievable phase sensitivities.

Using a balanced homodyne detection technique and various Gaussian input states, we show that far from being unphysical, the QCRB implied by the single parameter QFI is actually meaningful for each and every considered input state. We find that a coherent plus squeezed vacuum input state can benefit from the availability of an external phase reference for a low squeezing factor and a high coherent amplitude  if a properly unbalanced interferometer is used. The restriction on the squeezing factor(s) disappears for a squeezed-coherent plus squeezed vacuum input, this state being probably the most interesting candidate to demonstrate the sizable enhancement that can be obtained by using an unbalanced interferometer and an external phase reference. 

We conclude that when assessing the ``resources that are actually not available'' one must carefully ponder the actual experimental setup. If an external phase reference is possible (through e. g. homodyne detection), then the single parameter quantum Fisher information might give the pertinent answer regarding the best possible phase sensitivity.

\begin{acknowledgments}

This work has been supported by the Extreme Light Infrastructure Nuclear Physics (ELI-NP) Phase II, a project co-financed by the Romanian Government and the European Union through the European Regional Development Fund and the Competitiveness Operational Programme (1/07.07.2016, COP, ID 1334).

\end{acknowledgments}


\appendix

\section{Two parameter Fisher information}
Using the field operator transformations  \eqref{eq:field_op_transf_MZI_a_hat} we find the (photon) number operator $\hat{n}_3=\hat{a}_3^\dagger\hat{a}_3$, namely
\begin{equation}
\label{eq:a3_dagger_a3_versus_a1_a0}
\hat{n}_3=\vert{R}\vert^2\hat{a}_0^\dagger\hat{a}_0
+\vert{T}\vert^2{\hat{a}_1^\dagger\hat{a}_1}
-T^*R\left({\hat{a}_0^\dagger}{\hat{a}_1}
-{\hat{a}_0}{\hat{a}_1^\dagger}\right)
\end{equation}
and similarly $\hat{n}_2$ can be deduced. Starting from equation \eqref{eq:Fisher_matrix_elements} and using the field operator transformations \eqref{eq:field_op_transf_MZI_a_hat}, after some calculations we arrive at the expression:
\begin{widetext}
\begin{eqnarray}
\label{eq:F_dd_FINAL_FORM_GENERAL_compact}
\mathcal{F}_{dd}=\left(\vert{T}\vert^2-\vert{R}\vert^2\right)^2
\left(
\Delta^2\hat{n}_0
+\Delta^2\hat{n}_1
\right)
+8\vert{TR}\vert^2
\left(
\langle\hat{n}_0\rangle\langle\hat{n}_1\rangle
-\vert\langle\hat{a}_0\rangle\vert^2
\vert\langle\hat{a}_1\rangle\vert^2
-\Re\left\{
\langle{(\hat{a}_0^\dagger)^2}\rangle\langle{\hat{a}_1^2}\rangle
-\langle{\hat{a}_0^\dagger}\rangle^2\langle{\hat{a}_1}\rangle^2
\right\}
\right)
\nonumber\\
+4\vert{TR}\vert^2
\left(
\langle\hat{n}_1\rangle
+\langle\hat{n}_0\rangle
\right)
-8|TR|\left(\vert{T}\vert^2-\vert{R}\vert^2\right)\left(
\Im\left\{\left(\langle{\hat{a}_0^\dagger\hat{n}_0}\rangle
-\langle{\hat{a}_0^\dagger}\rangle\langle{\hat{n}_0}\rangle\right)
\langle{\hat{a}_1}\rangle
+\langle{\hat{a}_0}\rangle\left(\langle{\hat{a}_1^\dagger\hat{n}_1}\rangle
-\langle{\hat{n}_1}\rangle\langle{\hat{a}_1^\dagger}\rangle\right)
\right\}
\right).
\end{eqnarray}
In the balanced case $\mathcal{F}_{dd}$ simplifies and the result is given by equation \eqref{eq:app:F_dd_FINAL_FORM_GENERAL_balanced}. The Fisher matrix term $\mathcal{F}_{sd}$ is found to be
\begin{eqnarray}
\label{eq:F_sd_GENERIC_FINAL_compact}
\mathcal{F}_{sd}
=\left(\vert{T}\vert^2-\vert{R}\vert^2\right)
\left(\Delta^2{\hat{n}_0}
-\Delta^2{\hat{n}_1}
\right)
+4|TR|\Im\left\{\langle{\hat{a}_0}\rangle\langle{\hat{a}_1^\dagger}\rangle
+\left(\langle{\hat{n}_0\hat{a}_0}\rangle
-\langle{\hat{n}_0}\rangle\langle{\hat{a}_0}\rangle\right)\langle{\hat{a}_1^\dagger}\rangle
+\langle{\hat{a}_0}\rangle
\left(\langle{\hat{a}_1^\dagger\hat{n}_1}\rangle
-\langle{\hat{a}_1^\dagger}\rangle
\langle{\hat{n}_1}\rangle\right)
\right\}.
\quad\quad
\end{eqnarray}

\section{Single parameter Fisher information $\mathcal{F}^{(i)}$}
\label{sec:app:single_Fisher_information_asym}
Starting from definition \eqref{eq:Fisher_information_single_i} and using the field operator transformations \eqref{eq:field_op_transf_MZI_a_hat}, after a number of calculations we arrive at the final result:
\begin{eqnarray}
\label{eq:Fisher_sg_param_nonBal_symBS_separable_input_GENERAL}
\mathcal{F}^{(i)}
=4\vert{R}\vert^4\Delta^2{\hat{n}_0}
+4\vert{T}\vert^4\Delta^2{\hat{n}_1}
+4\vert{TR}\vert^2\left(
\langle{\hat{n}_0}\rangle
+\langle{\hat{n}_1}\rangle
+2(\langle{\hat{n}_0}\rangle\langle{\hat{n}_1}\rangle
-\vert\langle{\hat{a}_0}\rangle\vert^2\vert\langle{\hat{a}_1}\rangle\vert^2)
\right)
\nonumber\\
-8\vert{TR}\vert^2\Re\left\{\langle{\hat{a}_0^2}\rangle\langle{(\hat{a}_1^\dagger)^2}\rangle
-\langle{\hat{a}_0}\rangle\langle{\hat{a}_1^\dagger}\rangle^2\right\}
-8|TR|\Im\left\{\langle{\hat{a}_0}\rangle\langle{\hat{a}_1^\dagger}\rangle
\right\}
\nonumber\\
-16|TR|\vert{R}\vert^2\Im\left\{\left(\langle{\hat{n}_0\hat{a}_0}\rangle
-\langle{\hat{n}_0}\rangle\langle{\hat{a}_0}\rangle\right)\langle{\hat{a}_1^\dagger}\rangle\right\}
-16|TR|\vert{T}\vert^2\Im\left\{\langle{\hat{a}_0}\rangle\left(\langle{\hat{a}_1^\dagger\hat{n}_1}\rangle
-\langle{\hat{n}_1}\rangle\langle{\hat{a}_1^\dagger}\rangle\right)\right\}.
\end{eqnarray}
In the balanced case, $\mathcal{F}^{(i)}$ reduces to
\begin{eqnarray}
\label{eq:Fisher_sg_param_nonBal_symBS_separable_input_balanced}
\mathcal{F}^{(i)}
=\Delta^2{\hat{n}_0}+\Delta^2{\hat{n}_1}
+\langle{\hat{n}_0}\rangle
+\langle{\hat{n}_1}\rangle
+2(\langle{\hat{n}_0}\rangle\langle{\hat{n}_1}\rangle
-\vert\langle{\hat{a}_0}\rangle\vert^2\vert\langle{\hat{a}_1}\rangle\vert^2)
-2\Re\left\{\langle{\hat{a}_0^2}\rangle\langle{(\hat{a}_1^\dagger)^2}\rangle
-\langle{\hat{a}_0}\rangle\langle{\hat{a}_1^\dagger}\rangle^2\right\}
\nonumber\\
-4\Im\left\{\langle{\hat{a}_0}\rangle\langle{\hat{a}_1^\dagger}\rangle
+\left(\langle{\hat{n}_0\hat{a}_0}\rangle
-\langle{\hat{n}_0}\rangle\langle{\hat{a}_0}\rangle\right)\langle{\hat{a}_1^\dagger}\rangle
+\langle{\hat{a}_0}\rangle\left(\langle{\hat{a}_1^\dagger\hat{n}_1}\rangle
-\langle{\hat{n}_1}\rangle\langle{\hat{a}_1^\dagger}\rangle\right)\right\}.
\end{eqnarray}

\section{Single parameter Fisher information $\mathcal{F}^{(ii)}$}
\label{sec:app:single_Fisher_information_sym}
From definition \eqref{eq:Fisher_information_single_ii}, using the field operator transformations \eqref{eq:field_op_transf_MZI_a_hat} we arrive at
\begin{eqnarray}
\label{eq:Fisher_information_sg_param_symmetrical_varphi_over_2}
\mathcal{F}^{(ii)}=
(\vert{T}\vert^4+\vert{R}\vert^4)(\Delta^2{\hat{n}_0}+\Delta^2{\hat{n}_1})
+2\vert{TR}\vert^2\left(
\langle{\hat{n}_0}\rangle
+\langle{\hat{n}_1}\rangle+2(\langle{\hat{n}_0}\rangle\langle{\hat{n}_1}\rangle
-\vert\langle{\hat{a}_0}\rangle\vert^2\vert\langle{\hat{a}_1}\rangle\vert^2)\right)
\nonumber\\
-2\vert{TR}\vert^2\left(\langle{\hat{a}_0^2}\rangle\langle{(\hat{a}_1^\dagger)^2}\rangle
+\langle{(\hat{a}_0^\dagger)^2}\rangle\langle{\hat{a}_1^2}\rangle
-\langle{\hat{a}_0}\rangle^2\langle{\hat{a}_1^\dagger}\rangle^2
-\langle{\hat{a}_0^\dagger}\rangle^2\langle{\hat{a}_1}\rangle^2
\right)
\nonumber\\
+2T^*R(\vert{T}\vert^2-\vert{R}\vert^2)\left(\langle{\hat{a}_0^\dagger}{\hat{n}_0}\rangle-\langle{\hat{a}_0^\dagger}\rangle\langle{\hat{n}_0}\rangle\right)\langle{\hat{a}_1}\rangle
-2T^*R(\vert{T}\vert^2-\vert{R}\vert^2)\left(\langle{\hat{n}_0}{\hat{a}_0}\rangle-\langle{\hat{n}_0}\rangle\langle{\hat{a}_0}\rangle\right)\langle{\hat{a}_1^\dagger}\rangle
\nonumber\\
+2T^*R(\vert{T}\vert^2-\vert{R}\vert^2)\langle{\hat{a}_0}\rangle\left(\langle{\hat{a}_1^\dagger\hat{n}_1}\rangle-\langle{\hat{a}_1^\dagger}\rangle\langle{\hat{n}_1}\rangle\right)
-2T^*R(\vert{T}\vert^2-\vert{R}\vert^2)\langle{\hat{a}_0^\dagger}\rangle\left(\langle{\hat{n}_1\hat{a}_1}\rangle-\langle{\hat{n}_1}\rangle\langle{\hat{a}_1}\rangle\right).
\end{eqnarray}
In the balanced case $\mathcal{F}^{(ii)}$ simplifies to
\begin{eqnarray}
\label{eq:Fisher_information_sg_param_symmetrical_varphi_over_2_BALANCED}
\mathcal{F}^{(ii)}=\frac{1}{2}\left(
\Delta^2{\hat{n}_0}+\Delta^2{\hat{n}_1}
+\langle{\hat{n}_0}\rangle
+\langle{\hat{n}_1}\rangle+2(\langle{\hat{n}_0}\rangle\langle{\hat{n}_1}\rangle
-\vert\langle{\hat{a}_0}\rangle\vert^2\vert\langle{\hat{a}_1}\rangle\vert^2)
-2\Re\left\{\langle{\hat{a}_0^2}\rangle\langle{(\hat{a}_1^\dagger)^2}\rangle
-\langle{\hat{a}_0}\rangle^2\langle{\hat{a}_1^\dagger}\rangle^2
\right\}
\right).
\end{eqnarray}
 
\section{The $\Upsilon$ functions}
\label{sec:app:Upsilon_functions}
We define the functions
\begin{equation}
\label{eq:function:Upsilon_Plus_Minus_Definition}
\Upsilon^{+/-}\left(\gamma,\chi\right)={\vert\gamma\vert^2}\left(\cosh2s
\pm\sinh2s\cos\left(2\theta_\gamma-\vartheta\right)\right)
\end{equation}
with both arguments complex, ${\gamma=\vert\gamma\vert e^{i\theta_\gamma}}$ and ${\chi=se^{i\vartheta}}$ with $s\in\mathbb{R}^+$, $\theta_\gamma,\vartheta\in[0,2\pi]$. These functions allow the compact writing of Fisher matrix coefficients as well as output variances for a range of Gaussian input states \cite{Ata19}. For the PMC $2\theta_\gamma-\vartheta=0$ we find $\Upsilon^{+}\left(\gamma,\chi\right)=\vert\gamma\vert^2e^{2s}$ and $\Upsilon^{-}\left(\gamma,\chi\right)=\vert\gamma\vert^2e^{-2s}$. For the PMC $2\theta_\gamma-\vartheta=\pm\pi$ we find $\Upsilon^{+}\left(\gamma,\chi\right)=\vert\gamma\vert^2e^{-2s}$ and $\Upsilon^{-}\left(\gamma,\chi\right)=\vert\gamma\vert^2e^{2s}$. See also Fig.~2 in reference \cite{Ata19}.

\section{QFI calculations for a double coherent input} 
\label{sec:app:Fisher_dual_coh_input}
From equation \eqref{eq:F_2p_definition} and using the results from equations \eqref{eq:F_dd_dual_coh} and \eqref{eq:F_sd_dual_coh} we get
\begin{eqnarray}
\label{eq:Fisher_2p_dual_coherent}
\mathcal{F}^{(2p)}
=4\vert{TR}\vert^2\left(\vert\alpha\vert^2+\vert\beta\vert^2\right)
-16\vert{TR}\vert^2\frac{\vert\alpha\beta\vert^2\sin^2\Delta\theta}{\vert\alpha\vert^2+\vert\beta\vert^2}
+4\left(\vert{T}\vert^2-\vert{R}\vert^2\right)^2\frac{\vert\alpha\beta\vert^2}{\vert\alpha\vert^2
+\vert\beta\vert^2}
\nonumber\\
-8\vert{TR}\vert\left(\vert{T}\vert^2-\vert{R}\vert^2\right)\vert\alpha\beta\vert\frac{\vert\alpha\vert^2-\vert\beta\vert^2}{\vert\alpha\vert^2+\vert\beta\vert^2}\sin\Delta\theta
.
\end{eqnarray}
In the asymmetric single phase scenario from Fig.~\ref{fig:MZI_Fisher_single_phase}, an optimum transmission coefficient for $BS_1$ (in the sense that it maximizes $\mathcal{F}^{(i)}$) can be found,
\begin{equation}
\label{eq:T_opt_squared_versus_Delta_theta_dual_coh_F_i}
T_{opt}^{(i)}=\frac{1}{\sqrt{2+\frac{(|\alpha|^2-|\beta|^2)^2}{2|\alpha\beta|^2\sin^2\Delta\theta}
-\frac{||\alpha|^2-|\beta|^2|}{|\alpha\beta||\sin\Delta\theta|}\sqrt{1+\frac{(|\alpha|^2-|\beta|^2)^2}{4|\alpha\beta|^2\sin^2\Delta\theta}}}
}
.
\end{equation}
 
\section{QFI calculations for a coherent plus squeezed vacuum input} 
\label{sec:app:Fisher_coh_plus_sqz_vac}
The QFI $\mathcal{F}^{(i)}$ from both equations \eqref{eq:F_i_coh_sqz_vac} and \eqref{eq:F_i_sqz-coh_plus_SQZ_VAC} can be put in the form $\mathcal{F}^{i}=A_f|T|^4+B_f|R|^4+C_f|TR|^2$, i. e.
\begin{equation}
\label{eq:F_i_Fisher_AfBfCf}
\mathcal{F}^{i}
=(A_f+B_f-C_f)T^4+(C_f-2B_f)T^2+B_f
\end{equation}
and without loss of generality, starting from equation \eqref{eq:F_i_Fisher_AfBfCf} we assume $T$ real. Differentiating with respect to $T^2$ and solving this equation brings us to
\begin{equation}
\label{eq:T_opt_GENERIC_Af_Bf_C_f}
T^{(i)}_{opt}=\sqrt{\frac{C_f-2B_f}{2(C_f-A_f+B_f)}}.
\end{equation}
For the input state \eqref{eq:psi_in_coherent_plus_squeezed_vacuum} we have the coefficients
\begin{equation}
\left\{
\begin{array}{l}
A_f=\vert\alpha\vert^2\\
B_f=\frac{\sinh^22r}{2}\\
C_f=\sinh^2r+\Upsilon^+\left(\alpha,\xi\right)
\end{array}
\right.
\end{equation}
and we arrive at the expression given by equation \eqref{eq:T_opt_coh_sqz_vac}. If $T^{(i)}_{opt}$ exists, replacing \eqref{eq:T_opt_coh_sqz_vac} into equation  \eqref{eq:F_i_coh_sqz_vac} yields the maximum single-parameter QFI 
\begin{eqnarray}
\label{eq:F_i_coh_sqz_vac_MAX}
\mathcal{F}_{max}^{(i)}=
\frac{(\Upsilon^+\left(\alpha,\xi\right))^2
+\sinh^4r\Upsilon^+\left(\alpha,\xi\right)
-2\sinh^22r\vert\alpha\vert^2
+\sinh^4r
}{\Upsilon^+\left(\alpha,\xi\right)-\vert\alpha\vert^2-\sinh^2r\cosh2r}
.
\end{eqnarray}
When discussing the conditions of existence of $0\leq T_{opt}^{(i)}\leq1$  one must note that \eqref{eq:T_opt_coh_sqz_vac} becomes meaningless when $r\to0$ (in this limit, equation \eqref{eq:F_i_Fisher_AfBfCf} actually degenerates to $\mathcal{F}^{(i)}=T^2\vert\alpha\vert^2$). In the following we assume PMC \eqref{eq:PMC_coh_sqz_vac} satisfied. We first define the limits:
\begin{equation}
\left\{
\begin{array}{c}
\alpha_{lim1}^2=\frac{(\cosh{2r}-1)(\cosh{2r}+0.5)}{e^{2r}}\\
\alpha_{lim2}^2=\frac{\sinh^2r}{|2-e^{2r}|}
\end{array}
\right.
\end{equation}
For small values of $r$ we have $\alpha_{lim2}^2<\alpha_{lim1}^2$. Thus, $T_{opt}^{(i)}$ exists if $\vert\alpha\vert^2\in[\alpha_{lim2}^2,\: \alpha_{lim1}^2]$. If $\alpha_{lim1}^2<\alpha_{lim2}^2$ and moreover $r\leq\ln2/2$ then $T_{opt}^{(i)}$ exists if $\vert\alpha\vert^2\in[\alpha_{lim1}^2,\: \alpha_{lim2}^2]$. Finally if $r>\ln2/2$, then $T_{opt}^{(i)}$ exists for $\vert\alpha\vert^2\geq\alpha_{lim1}^2$.

\section{QFI calculations for a squeezed-coherent plus squeezed vacuum input} 
\label{sec:app:Fisher_sqz_coh_plus_sqz_vac}
For a squeezed-coherent state at input port $1$ we have 
$\Delta^2\langle\hat{n}_1\rangle
=\frac{\sinh^22z}{2}+\Upsilon^-\left(\alpha,\zeta\right)$
\cite{Ata19} and we employed this result in computing $\mathcal{F}_{ss}$ from equation \eqref{eq:F_ss_sqz_coh_plus_sqz_vac_symm_BS_non_balanced}. Using the input state \eqref{eq:psi_in_sqz-coh_sqz_vac} and the definition of the Fisher matrix element $\mathcal{F}_{dd}$ we get
\begin{eqnarray}
\label{eq:F_dd_coh_sqz_vac_SQZ_VAC}
\mathcal{F}_{dd}=\left(\vert{T}\vert^2-\vert{R}\vert^2\right)^2
\bigg(\frac{\sinh^2{{2r}}}{2}
+\frac{\sinh^2{{2z}}}{2}
+\Upsilon^-\left({\alpha,\zeta}\right)\bigg)
+4\vert{TR}\vert^2
\big(\Upsilon^+\left(\alpha,\xi\right)
+\sinh^2{r}
+\sinh^2{z}
\nonumber\\
+2\sinh{r}\sinh{z}
\left(\sinh{r}\sinh{z}
-\cosh{r}\cosh{z}\cos(\phi-\theta)\right)
\big)
\end{eqnarray}
Finally, starting from equation \eqref{eq:F_sd_GENERIC_FINAL_compact},
$\mathcal{F}_{sd}$ is found to be
\begin{eqnarray}
\label{eq:F_sd_coh_sqz_vac_SQZ_VAC2}
\mathcal{F}_{sd}
=\left(\vert{T}\vert^2-\vert{R}\vert^2\right)
\left(\frac{\sinh^2{2r}}{2}
-\frac{\sinh^2{2z}}{2}
-\Upsilon^-\left({\alpha,\zeta}\right)
\right)
\end{eqnarray}
From definition \eqref{eq:F_2p_definition} and the previous results, we get the two-parameter QFI,
\begin{eqnarray}
\label{eq:F2p_sqz-coh_sqz_vac_symm_BS_non_balanced}
\mathcal{F}^{(2p)}=4\vert{TR}\vert^2
\left(\Upsilon^+\left(\alpha,\xi\right)
+\frac{\cosh{2r}\cosh{2z}
-\sinh{2r}\sinh{2z}\cos(\phi-\theta)-1}{2}
\right)
\nonumber\\
+\left(\vert{T}\vert^2-\vert{R}\vert^2\right)^2\frac{\sinh^2{2r}\left(\sinh^2{2z}+2\Upsilon^-\left({\alpha,\zeta}\right)\right)}
{\frac{\sinh^2{2r}}{2}+\frac{\sinh^22z}{2}+\Upsilon^-\left({\alpha,\zeta}\right)}
.
\end{eqnarray} 
For the asymmetric phase shift case form Fig.~\ref{fig:MZI_Fisher_single_phase}, the single-parameter QFI yields
\begin{eqnarray}
\label{eq:F_i_sqz-coh_plus_SQZ_VAC}
\mathcal{F}^{(i)}=
4|T|^4\left(\frac{\sinh^2{{2z}}}{2}
+\Upsilon^-\left({\alpha,\zeta}\right)\right)
+4|R|^4\frac{\sinh^2{{2r}}}{2}
\nonumber\\
+4\vert{TR}\vert^2
\left(\Upsilon^+\left(\alpha,\xi\right)
+\sinh^2{r}
+\sinh^2{z}
+2\sinh{r}\sinh{z}
\left(\sinh{r}\sinh{z}
-\cosh{r}\cosh{z}\cos(\phi-\theta)\right)
\right)
.
\end{eqnarray}
If an optimal transmission factor $0<T^{(i)}_{opt}<1$ exists (in the sense that it maximizes $\mathcal{F}^{(i)}$), then it is given by equation \eqref{eq:T_opt_GENERIC_Af_Bf_C_f} with the coefficients
\begin{equation}
\label{eq:T_opt_coefficients_Af_Bf_C_f_sqzcoh_plus_sqzvac}
\left\{
\begin{array}{l}
A_f=\frac{\sinh^22z}{2}+\Upsilon^-(\alpha,\zeta)\\
B_f=\frac{\sinh^22r}{2}\\
C_f=\Upsilon^+\left(\alpha,\xi\right)
+\sinh^2{r}+\sinh^2{z}+2\sinh{r}\sinh{\blue z}
(\sinh{r}\sinh{z}
-\cosh{r}\cosh{z}\cos(\phi-\theta))
\end{array}
\right.
\end{equation}
In Section \ref{subsec:Fisher_coh_sqz_vac} we concluded that the $\Upsilon^+\left(\alpha,\xi\right)$ term dominates all other terms from equations \eqref{eq:Fisher_2p_coh_sqz_vac} and \eqref{eq:F_i_coh_sqz_vac} in the high-$\alpha$ regime. This assertion is still true for $\mathcal{F}^{(2p)}$ from equation \eqref{eq:F2p_sqz-coh_sqz_vac_symm_BS_non_balanced}. But, as mentioned in Section \ref{subsec:Fisher_sqz-coh_sqz_vac}, in the high-$\alpha$ regime $\mathcal{F}^{(i)}$ does not necessarily maximize in the balanced case. Indeed, $\mathcal{F}^{(i)}$ from equation \eqref{eq:F_i_sqz-coh_plus_SQZ_VAC} approximates in this regime to
\begin{equation}
\mathcal{F}^{(i)}\sim4|T|^4\Upsilon^-\left(\alpha,\zeta\right)+4|TR|^2\Upsilon^+\left(\alpha,\xi\right).
\end{equation}
and we immediately find the optimum transmission coefficient,
\begin{equation}
\label{eq:app:T_opt_i_sqzcoh_sqzvac_high_alpha}
T^{(i)}_{opt}\approx
\sqrt{\frac{\Upsilon^+\left(\alpha,\xi\right)}{2\vert\Upsilon^+\left(\alpha,\xi\right)-\Upsilon^-(\alpha,\zeta)\vert}}
\end{equation}
For the optimum PMCs \eqref{eq:PMC_sqz-coh_plus_sqz-vac} satisfied, we arrive at $T^{(i)}_{opt}$ from equation \eqref{eq:T_opt_i_sqzcoh_sqzvac_high_alpha}. If, from equation \eqref{eq:T_opt_i_sqzcoh_sqzvac_high_alpha} one obtains $T^{(i)}_{opt}>1$, this simply implies that the optimum transmission factor for $BS_1$ is $T^{(i)}_{opt}=1$.

For the symmetric phase shift scenario we have the QFI
\begin{eqnarray}
\label{eq:F_ii_sqz-coh_plus_SQZ_VAC}
\mathcal{F}^{(ii)}=\frac{\sinh^22r}{2}
+\frac{\sinh^22z}{2}
+\Upsilon^-\left(\alpha,\zeta\right)
+2\vert{TR}\vert^2
\bigg(\Upsilon^+\left(\alpha,\xi\right)
-\Upsilon^-\left({\alpha,\zeta}\right)
\nonumber\\
-\sinh^2{r}\cosh2r
-\sinh^2{z}\cosh2z
+2\sinh{r}\sinh{z}
\left(\sinh{r}\sinh{z}
-\cosh{r}\cosh{z}\cos(\phi-\theta)\right)
\bigg)
\end{eqnarray}
and if the condition 
\begin{equation}
\label{eq:Sqz-coh_plus_SQZ_VAC_CONDITION_F_ii_max_bal}
\Upsilon^+\left(\alpha,\xi\right)
-\Upsilon^-\left({\alpha,\zeta}\right)
-\sinh^2{r}\cosh2r
-\sinh^2{z}\cosh2z
+2\sinh{r}\sinh{z}
\left(\sinh{r}\sinh{z}
-\cosh{r}\cosh{z}\cos(\phi-\theta)\right)>0
\end{equation}
is satisfied then $\mathcal{F}^{(ii)}$ maximizes in the balanced case.
 
\section{Difference-intensity detection}
\label{sec:app:MZI_both_unbalanced_diff_det}
From Fig.~\ref{fig:Figure1_MZI_2D_differential_homodyne_v01}, using the field operator transformations \eqref{eq:field_op_transf_MZI_a_hat} and
\begin{equation}
\label{eq:field_op_transf_BS2_unbalanced}
\left\{
\begin{array}{l}
\hat{a}_5=R'{\hat{a}'_2}+T'{\hat{a}'_3}\\
\hat{a}_4=T'{\hat{a}'_2}+R'{\hat{a}'_3}
\end{array}
\right.
\end{equation}
where we recall that $T'$ ($R'$) denote the transmission (reflection) coefficients of $BS_2$,
we can write the field operator transformations
\begin{equation}
\label{eq:field_op_transf_NON_balanced_MZI_sym_BS_bothBS_unbal}
\left\{
\begin{array}{l}
\hat{a}_4=e^{-i\varphi_2}\left[\left(TT'+RR'e^{-i\varphi}\right){\hat{a}_0}
+\left(TR'e^{-i\varphi}+RT'\right){\hat{a}_1}\right]\\
{\mbox{ }}\\
\hat{a}_5=e^{-i\varphi_2}\left[\left(TR'+RT'e^{-i\varphi}\right){\hat{a}_0}
+\left(TT'e^{-i\varphi}+RR'\right){\hat{a}_1}\right]\\
\end{array}
\right.
\end{equation}
where $\varphi=\varphi_1-\varphi_2$. The final expression for $\hat{N}_d$ is
\begin{eqnarray}
\label{eq:Nd_both_BS_unbalanced}
\hat{N}_d=\left((|T|^2-|R|^2)(|T'|^2-|R'|^2)
-4|TRT'R'|\cos\varphi\right)\left({\hat{n}_0}-{\hat{n}_1}\right)
\nonumber\\
+2\left(T^*R(|R'|^2-|T'|^2)+(|R|^2e^{-i\varphi}-|T|^2e^{i\varphi}){T'}^*R'\right){\hat{a}_0}{\hat{a}_1^\dagger}
\nonumber\\
+2\left(T^*R(|T'|^2-|R'|^2)
+(|T|^2e^{-i\varphi}
-|R|^2e^{i\varphi}){T'}^*R'\right){\hat{a}_0^\dagger}{\hat{a}_1}
\end{eqnarray}
One can see from equation \eqref{eq:Nd_both_BS_unbalanced} that $\hat{N}_d$ depends only on $\varphi=\varphi_1-\varphi_2$ thus insensitive to an external (or global) phase. The derivative of $\langle\hat{N}_d\rangle$ with respect to $\varphi$ yields
\begin{eqnarray}
\label{eq:Nd_avg_del_varphi_both_BS_unbalanced}
\frac{\partial\langle\hat{N}_d\rangle}{\partial\varphi}=
4\left(|TR|\sin\varphi(\langle{\hat{n}_0}\rangle-\langle{\hat{n}_1}\rangle)
+\Re\left\{(|R|^2e^{-i\varphi}+|T|^2e^{i\varphi})\langle{\hat{a}_0}\rangle\langle{\hat{a}_1^\dagger}\rangle\right\}\right)|T'R'|
\end{eqnarray}
For the variance we find 
\begin{eqnarray}
\label{eq:Variance_Nd_both_BS_unbalanced}
\Delta\hat{N}_d=
A_d^2(\Delta^2{\hat{n}_0}
+\Delta^2{\hat{n}_1})
+|C_d|^2(\langle{\hat{n}_0}\rangle
+\langle{\hat{n}_1}\rangle)
+2|C_d|^2(\langle{\hat{n}_0}\rangle\langle{\hat{n}_1}\rangle
-\vert\langle{\hat{a}_0}\rangle\vert^2\vert\langle{\hat{a}_1}\rangle\vert^2)
\nonumber\\
+2\Re\left\{C_d^2(\langle{\hat{a}_0^2}\rangle\langle{(\hat{a}_1^\dagger)^2}\rangle
-\langle{\hat{a}_0}\rangle^2\langle{\hat{a}_1^\dagger}\rangle^2)\right\}
+4A_d\Re\left\{C_d\left((\langle{\hat{n}_0}{\hat{a}_0}\rangle
-\langle{\hat{n}_0}\rangle\langle{\hat{a}_0}\rangle)\langle{\hat{a}_1^\dagger}\rangle
-\langle{\hat{a}_0}\rangle(\langle{\hat{a}_1^\dagger}{\hat{n}_1}\rangle
-\langle{\hat{n}_1}\rangle\langle{\hat{a}_1^\dagger}\rangle)\right)
\right\}
\end{eqnarray}
where we made the notations
\begin{equation}
\label{eq:Ad_C_d_coefficients_diff_det_DEFINITION}
\left\{
\begin{array}{l}
A_d=1-2(|T||R'|+|R||T'|)^2
+4|TR||T'R'|(1-\cos\varphi)\\
C_d=2|T'R'|\sin\varphi+2i\left(|TR|(|R'|^2-|T'|^2)+(1-2|T|^2)|T'R'|\cos\varphi\right)
\end{array}
\right.
\end{equation}
and by direct calculation we also find the constraint
\begin{equation}
\label{eq:A_s_squared_plus_Cd_modsquare_is_1}
A_d^2+|C_d|^2=1.
\end{equation}
 
\section{Balanced homodyne detection}
\label{sec:app:homodyne}
From equations \eqref{eq:field_op_transf_NON_balanced_MZI_sym_BS_bothBS_unbal} and using the definition of  $\hat{X}_{\phi_L}$ we have 
\begin{eqnarray}
\label{eq:X_phi_L_homodyne_non_bal_sym_BS_GENERAL}
\langle\hat{X}_{\phi_L}\rangle=\Re\left\{e^{-i\phi_L}\left(\left(TT'e^{-i\varphi_2}+RR'e^{-i\varphi_1}\right)\langle{\hat{a}_0}\rangle
+\left(TR'e^{-i\varphi_1}+RT'e^{-i\varphi_2}\right)\langle{\hat{a}_1}\rangle\right)\right\}
\end{eqnarray}
For the asymmetric phase shift scenario from Fig.~\ref{fig:MZI_Fisher_single_phase} we have $\varphi_1=\varphi$ and $\varphi_2=0$. The derivative of $\langle\hat{X}_{\phi_L}\rangle$ with respect to $\varphi$ gives the expression from equation \eqref{eq:del_X_phi_L_avg__del_varphi_homodyne_non_bal_one_phase}. For the symmetric scenario from Fig.~\ref{fig:MZi_Fisher_single_param_sym_varphi_over_2} we have $\varphi_1=\varphi/2$ and $\varphi_2=-\varphi/2$, thus we get the result from equation
\eqref{eq:del_X_phi_L_avg__del_varphi_homodyne_non_bal_two_phases}. The variance of $\hat{X}_{\phi_L}$ is found to be
\begin{eqnarray}
\label{eq:variance_X_phi_L_homodyne}
\Delta^2\hat{X}_{\phi_L}=\frac{1}{4}+2\Re\left\{
A^2\Delta^2{\hat{a}_0}+B^2\Delta^2{\hat{a}_1}\right\}
+2|A|^2(\langle\hat{n}_0\rangle-\vert\langle{\hat{a}_0}\rangle\vert^2)
+2|B|^2(\langle{\hat{n}_1}\rangle-\vert\langle{\hat{a}_1}\rangle\vert^2)
\qquad
\end{eqnarray}
where we have the coefficients
\begin{equation}
\left\{
\begin{array}{l}
A=\frac{1}{2}e^{-i(\phi_L+\varphi_2)}\left(TT'+RR'e^{-i\varphi}\right)\\
B=\frac{1}{2}e^{-i(\phi_L+\varphi_2)}\left(TR'e^{-i\varphi}+T'R\right).
\end{array}
\right.
\end{equation}

\section{Phase sensitivity calculations for a dual coherent input}
\label{sec:app:phase_sensitivity_dual_coh}
For difference intensity detection scheme and a dual coherent input we get
\begin{eqnarray}
\label{eq:Nd_avg_del_varphi_both_BS_unbalanced_DUAL_COH_general}
\bigg\vert\frac{\partial\langle\hat{N}_d\rangle}{\partial\varphi}\bigg\vert=
4\big\vert|TR|\sin\varphi(\beta\vert^2-\vert\alpha\vert^2)
+\vert\alpha\beta\vert\left(|R|^2\cos(\theta_\alpha-\theta_\beta+\varphi)+|T|^2\cos(\theta_\alpha-\theta_\beta-\varphi\right)\big\vert\vert T'R'\vert
\end{eqnarray}
and the variance is simply $\Delta^2\hat{N}_d=\vert\alpha\vert^2+\vert\beta\vert^2$. The phase sensitivity is optimized for the input PMC {$\Delta\theta=0$} and equation \eqref{eq:Nd_avg_del_varphi_both_BS_unbalanced_DUAL_COH_general} becomes
\begin{equation}
\bigg\vert\frac{\partial\langle\hat{N}_d\rangle}{\partial\varphi}\bigg\vert=
4\big\vert|TR|\sin\varphi(\beta\vert^2-\vert\alpha\vert^2)
+\vert\alpha\beta\vert\cos\varphi\big\vert\vert T'R'\vert.
\end{equation}
For $T$ given, this expression is maximized at the optimum internal phase sift
\begin{equation}
\label{eq:varphi_OPT_dual_coh}
\varphi_{opt}
=\arccos\left(\frac{\vert\alpha\beta\vert}{\sqrt{\vert\alpha\beta\vert^2+|TR|^2(\beta\vert^2-\vert\alpha\vert^2)^2}}\right)
\end{equation}
yielding the phase sensitivity at the optimum angle
\begin{equation}
\Delta\tilde{\varphi}_{df}=\frac{\sqrt{\vert\alpha\vert^2+\beta\vert^2}}{4\sqrt{\vert\alpha\beta\vert^2(1-4|TR|^2)+|TR|^2(\vert\alpha\vert^2+\beta\vert^2)^2}\vert T'R'\vert}
.
\end{equation}
We can further optimize $\Delta\tilde{\varphi}_{df}$ by imposing both $BS_1$ and $BS_2$ balanced and we arrive at the expression \eqref{sec:Delta_varphi_df_dual_coh_OPTIMAL}.

\section{Phase sensitivity calculations for a coherent plus squeezed vacuum input}
\label{sec:app:phase_sensitivity_coh_sqz_vac}
For the variance of the output number operator we find the expression
\begin{eqnarray}
\label{eq:Variance_Nd_both_BS_unbalanced_coh_sqz_vac}
\Delta^2\hat{N}_d=
A_d^2\frac{\sinh^22r}{2}
+\vert\alpha\vert^2
+|C_d|^2(\sinh^2r
+2\vert\alpha\vert^2\sinh^2r)
-\sinh2r\vert\alpha\vert^2\Re\left\{C_d^2e^{-i(2\theta_\alpha-\theta)}\rangle
\right\}
\end{eqnarray} 
where the coefficients $A_d$ and $C_d$ were defined in equation \eqref{eq:Ad_C_d_coefficients_diff_det_DEFINITION}. The expression of $\Delta^2\hat{N}_d$ for the balanced case can be  found in the literature \cite{Dem15,API18}.

In case of a balanced homodyne detection, the variance $\Delta^2\hat{X}_{\phi_L}$ from equation \eqref{eq:variance_X_phi_L_homodyne} becomes
\begin{eqnarray}
\label{eq:variance_X_phi_L_homodyne_symbolic}
\Delta^2\hat{X}_{\phi_L}=\frac{1}{4}+2\Re\left\{
A^2\Delta^2{\hat{a}_0}\right\}
+2|A|^2\langle\hat{n}_0\rangle
\end{eqnarray}
and the phase sensitivity is found to be
\begin{equation}
\label{eq:Delta_varphi_hom_coh_sqz_vac_symbolic}
\Delta\varphi_{hom}^{(i)}=\frac{\sqrt{\frac{1}{4}
+2|A|^2\sinh^2r-\Re\left\{
A^2e^{i\theta}\right\}\sinh2r}}{2|TR'|\vert\alpha\vert\vert\cos\varphi\vert}
.
\end{equation}
Assuming  $\phi_L=\theta_\alpha$ and the PMCs from equation \eqref{eq:PMC_coh_sqz_vac} satisfied, the phase sensitivity can be written as
\begin{equation}
\label{eq:Delta_varphi_hom_coh_sqz_vac_PMC_optimal}
\Delta\varphi_{hom}^{(i)}=\frac{\sqrt{1
-2(|TT'|+|RR'|)^2\sinh re^{-r}
+|RR'|^2\sinh2r(1-\cos(2\varphi))
+4|TRT'R'|\sinh re^{-r}(1+\cos\varphi})}
{2|TR'|\vert\alpha\vert\vert\cos\varphi\vert}
.
\end{equation}

\section{Phase sensitivity calculations for a squeezed-coherent plus squeezed vacuum input}
Assuming $\phi_L=\theta_\alpha$ and the PMCs \eqref{eq:PMC_sqz-coh_plus_sqz-vac} satisfied, the variance of the operator $\hat{X}_L$ is found to be
\begin{eqnarray}
\label{eq:Variance_X_sqz_coh_squeezed_vacuum_FINAL}
\Delta^2\hat{X}_{\phi_L}=\frac{1}{4}
-\frac{1}{2}\left(|TT'|+|RR'|\right)^2\sinh^{r}e^{-r}
-\frac{1}{2}\left(|RT'|-|TR'|\right)^2\sinh{z}e^{-z}
\nonumber\\
+\frac{1}{2}|RR'|^2\sinh{r}\cosh{r}(1-\cos(2\varphi))
+|TRT'R'|\sinh{r}e^{-r}(1+\cos\varphi)
\nonumber\\
+\frac{1}{2}|TR'|^2\sinh{z}\cosh{z}(1-\cos(2\varphi))
-|TRT'R'|\sinh{z}e^{-z}(1+\cos\varphi).
\end{eqnarray}
If we now impose the optimum working point $\varphi_{opt}=(2k+1)\pi$ (with $k\in\mathbb{Z}$) we get the phase sensitivity at the optimal angle
\begin{equation}
\label{eq:Delta_varphi_sqz_coh_squeezed_vacuum_varphi_OPTIMAL}
\Delta\tilde{\varphi}_{hom}^{(i)}
=\frac{\sqrt{1-\left(|TT'|+|RR'|\right)^2(1-e^{-2r})
-\left(|RT'|-|TR'|\right)^2(1-e^{-2z})}}{2|TR'|\vert\alpha\vert}
\end{equation}
In the balanced case ($T=T'=1/\sqrt{2}$) this expression reduces to $\Delta\tilde{\varphi}_{hom}^{(i)}=e^{-r}/\vert\alpha\vert$, a result found in the literature \cite{Pre19,Ata19}. An optimum transmission coefficient for $BS_2$ (in the sense that with $T$ replaced by ${T}^{(i)}_{opt}$ it minimizes $\Delta\tilde{\varphi}_{hom}^{(i)}$) can be obtained from \eqref{eq:Delta_varphi_sqz_coh_squeezed_vacuum_varphi_OPTIMAL}, and we get
\begin{equation}
\label{eq:T_prime_opt_sqzcoh_plus_sqzvac}
{T'}^{(i)}_{opt}
=\frac{{T}^{(i)}_{opt}\sqrt{1-\left({T}^{(i)}_{opt}\right)^2}\vert e^{-2r}-e^{-2z}\vert}
{\sqrt{e^{-4z}
-\left({T}^{(i)}_{opt}\right)^2\left(e^{-4z}-e^{-4r}\right)}}
\end{equation}
where we recall that ${T}^{(i)}_{opt}$ is obtained from equation \eqref{eq:T_opt_GENERIC_Af_Bf_C_f} with the coefficients given by equation \eqref{eq:T_opt_coefficients_Af_Bf_C_f_sqzcoh_plus_sqzvac}.

\end{widetext}


\onecolumngrid


%
%

\bibliographystyle{apsrev4-1}

\bibliography{MZI_phase_sensitivity_bibtex}

\end{document}